\documentclass[sn-mathphys]{sn-jnl}
\usepackage{amssymb}
\usepackage{amsmath}
\usepackage{graphicx}
\usepackage[countmax]{subfloat}
\usepackage{float}
\usepackage{hyperref}
\usepackage{subfig}

\usepackage[charter,cal=cmcal]{mathdesign}
\usepackage{threeparttable}

\jyear{2025}%

\theoremstyle{thmstyleone}%
\theoremstyle{thmstyletwo}%

\theoremstyle{thmstylethree}%

\begin{document}

\title[Article Title]{HAQA: A Hardware-Guided and Fidelity-Aware Strategy for Efficient Qubit Mapping Optimization}


\author[1]{\fnm{Wenjie} \sur{Sun}}\email{202211022528@std.uestc.edu.cn}

\author*[2]{\fnm{Xiaoyu} \sur{Li}}\email{xiaoyuuestc@uestc.edu.cn}

\author[3]{\fnm{Lianhui} \sur{Yu}}\email{202222120303@std.uestc.edu.cn}

\author[1]{\fnm{Zhigang} \sur{Wang}}\email{zhigangwang@uestc.edu.cn}

\author[4]{\fnm{Geng} \sur{Chen}}\email{202312081614@std.uestc.edu.cn}

\author[5]{\fnm{Desheng} \sur{Zheng}}\email{zheng\_de\_sheng@163.com}

\author[4]{\fnm{Guowu} \sur{Yang}}\email{guowu@uestc.edu.cn}

\affil[1]{\orgdiv{The School of Electronic Science and Engineering}, \orgname{University of Electronic Science and Technology of China}, \orgaddress{\street{No.2006, Xiyuan Ave}, \city{Chengdu}, \postcode{611731}, \state{Sichuan}, \country{China}}}

\affil*[2]{\orgdiv{The School of Information and Software Engineering}, \orgname{University of Electronic Science and Technology of China}, \orgaddress{\street{No.2006, Xiyuan Ave}, \city{Chengdu}, \postcode{611731}, \state{Sichuan}, \country{China}}}

\affil[3]{\orgdiv{The School of Physics}, \orgname{University of Electronic Science and Technology of China}, \orgaddress{\street{No.2006, Xiyuan Ave}, \city{Chengdu}, \postcode{611731}, \state{Sichuan}, \country{China}}}

\affil[4]{\orgdiv{The School of Computer Science and Engineering}, \orgname{University of Electronic Science and Technology of China}, \orgaddress{\street{No.2006, Xiyuan Ave}, \city{Chengdu}, \postcode{611731}, \state{Sichuan}, \country{China}}}

\affil[5]{\orgdiv{The School of Computer Science and Software Engineering}, \orgname{Southwest Petroleum University}, \orgaddress{\street{No.8, Xindu Road}, \city{Chengdu}, \postcode{610500}, \state{Sichuan}, \country{China}}}


\abstract{
	Quantum algorithms rely on quantum computers for implementation, but the physical connectivity constraints of modern quantum processors impede the efficient realization of quantum algorithms. Qubit mapping, a critical technology for practical quantum computing applications, directly determines the execution efficiency and feasibility of algorithms on superconducting quantum processors. Existing mapping methods overlook intractable quantum hardware fidelity characteristics, reducing circuit execution quality. They also exhibit prolonged solving times or even failure to complete when handling large-scale quantum architectures, compromising efficiency. To address these challenges, we propose a novel qubit mapping method HAQA. HAQA first introduces a community-based iterative region identification strategy leveraging hardware connection topology, achieving effective dimensionality reduction of mapping space. This strategy avoids global search procedures, with complexity analysis demonstrating quadratic polynomial-level acceleration. Furthermore, HAQA implements a hardware-characteristic-based region evaluation mechanism, enabling quantitative selection of mapping regions based on fidelity metrics. This approach effectively integrates hardware fidelity information into the mapping process, enabling fidelity-aware qubit allocation. Experimental results demonstrate that HAQA significantly improves solving speed and fidelity while ensuring solution quality. When applied to state-of-the-art quantum mapping techniques Qsynth-v2 and TB-OLSQ2, HAQA achieves acceleration ratios of 632.76× and 286.87× respectively, while improving fidelity by up to 52.69\% and 238.28\%.
	
	}


\maketitle

\section{Introduction}\label{sec1}

Quantum computing, with its unique parallel processing capabilities and potential for efficiently solving complex problems, holds promise for groundbreaking advancements across various fields. With the rapid advancement of quantum hardware, quantum computers have significantly increased in scale in recent years. Leading companies such as Google \cite{Arute2019}, IBM \cite{JerryChow,author.fullNamevphantom,2024}, and Rigetti \cite{Rigetti&Co2024} have continuously developed quantum computers with increasingly larger qubit counts. The most advanced superconducting quantum computers now boast over $1000$ qubits \cite{author.fullNamevphantom}. Quantum compilation \cite{khatri2019quantum}, serving as a critical bridge between quantum algorithms and quantum hardware, plays a fundamental role in realizing quantum computing applications. The core concept of quantum compilation involves transforming high-level quantum algorithms into executable quantum circuits while considering various hardware constraints. These constraints include not only the architectural limitations of quantum computers but also the noise characteristics in the current Noisy Intermediate-Scale Quantum (NISQ) era \cite{Preskill2018}, making quantum compilation an essential step towards practical quantum computing.

As a core technology in quantum compilation, qubit mapping, alternatively termed quantum layout synthesis\cite{Tan2020}, is a crucial optimization process that involves deploying quantum circuits onto quantum computing hardware while navigating complex physical constraints. These constraints include physical qubit connectivity limitations\cite{Tan2020} and quantum gate fidelity variations\cite{Ferrari2022}. A effective qubit mapping strategies can substantially reduce quantum circuit depth and significantly improve circuit execution reliability\cite{Ferrari2022}. The fundamental objective of qubit mapping is to bridge the gap between the abstract quantum circuit design and the physical constraints of quantum hardware, ultimately enhancing computational performance and reliability.

Currently, qubit mapping methods can be categorized into two primary approaches: heuristic and solver-based methods. Heuristic methods leverage meta-heuristic algorithms for qubit mapping\cite{Li2019,Niu2020,Sivarajah2020}, with representative algorithms including SABRE proposed by Li\cite{Li2019}. While these approaches offer rapid solution generation, they yield solution quality that is inferior to solver-based methods. 
Solver-based methods reframe qubit mapping as a SAT problem, encoding solving contexts as constraints and employing incremental solving to optimize multiple circuit metrics such as circuit depth and swap gate number\cite{Wille2014,Wille2019,Tan2020,Tan2021,Lin2023,Shaik2023,Shaik2024}.The solver-based qubit mapping approach was first introduced by Robert Wille in 2014\cite{Wille2014}. Subsequent research saw significant methodological advancements: in 2020, Bochen Tan et al. proposed a two-stage search method improving circuit quality\cite{Tan2020}, and in 2021, they further reduced circuit depth using gate absorption techniques\cite{Tan2021}. In 2023, Wan-Hsuan Lin et al. enhanced method scalability through reduced redundant constraints and incremental solving\cite{Lin2023}. In the same year, Irfansha Shaik et al. introduced optimal classical planners for layout optimization\cite{Shaik2023}. In 2024, Shaik proposed SAT encoding based on parallel plans to improve scalability\cite{Shaik2024}. Despite these innovations, solver methods consistently face two fundamental challenges as the \textbf{\emph{main motivation}} for our work:

\begin{enumerate}
	
	\item{Solving efficiency dramatically decreases with increasing physical qubit count. As quantum architectures grow in scale, both the variable set and constraints in the solver expand substantially, leading to significantly prolonged solving time or even failure to converge.}
	\item{Current methods exclusively consider circuit depth and swap gate numbers while neglecting critical fidelity and quantum hardware physical characteristics. This limitation becomes increasingly significant as quantum computers evolve with more complex architectures and intricate distribution of gate fidelities.}
	
\end{enumerate}

These challenges substantially compromise the practical utility of solver-based approaches in real scenarios.

To address these challenges, we propose HAQA, a optimization method designed to enhance qubit mapping through hardware-oriented region identification mechanisms. Our method introduces a community-based iterative region identification strategy leveraging hardware connection topology, achieving effective dimensionality reduction of mapping space and avoiding global search procedures. From the solver's perspective, this process effectively prunes the solution space by utilizing hardware topology prior knowledge. Moreover, HAQA implements a hardware-characteristic-based region evaluation mechanism, enabling quantitative selection of mapping regions based on fidelity metrics and addressing the inherent inefficiencies of classical solvers in fidelity considerations. The \textbf{\emph{main contributions}} of our work as follow:

\begin{enumerate}
	
	\item{We introduce a hardware-aware and adaptive acceleration method compatible with diverse classical solver methodologies. The proposed method mitigates the efficiency bottleneck and complements fidelity optimization limitations in current solver-based qubit mapping approaches.}
	\item{We propose a novel technical framework that integrates hardware characteristics into the mapping process. The method leverages community-based iterative region identification for efficient solution space reduction, while implementing quantitative fidelity evaluation mechanisms for mapping region selection. This transforms the global mapping problem into a region guided optimization process. Additionally, a transferable complexity analysis framework is provided, demonstrating the method's polynomial-level acceleration potential.
	}
	\item{Experimental validation on two state-of-the-art solvers demonstrates significant performance improvements. The method achieves acceleration ratios of $632.76\times$ and $286.87\times$ for Qsynth-v2 and TB-OLSQ2 respectively, while concurrently improving fidelity by $52.69\%$ and $238.28\%$. These advantages become increasingly prominent as quantum computing architectures grow in scale and complexity.}
\end{enumerate}

\section{Preliminaries}
\subsection{Qubit mapping}
\begin{figure}[tb]
	\begin{minipage}[b]{\columnwidth}
		\centering
		\subfloat[Quantum circuit before qubit mapping.]
		{\includegraphics[width=0.27\columnwidth]{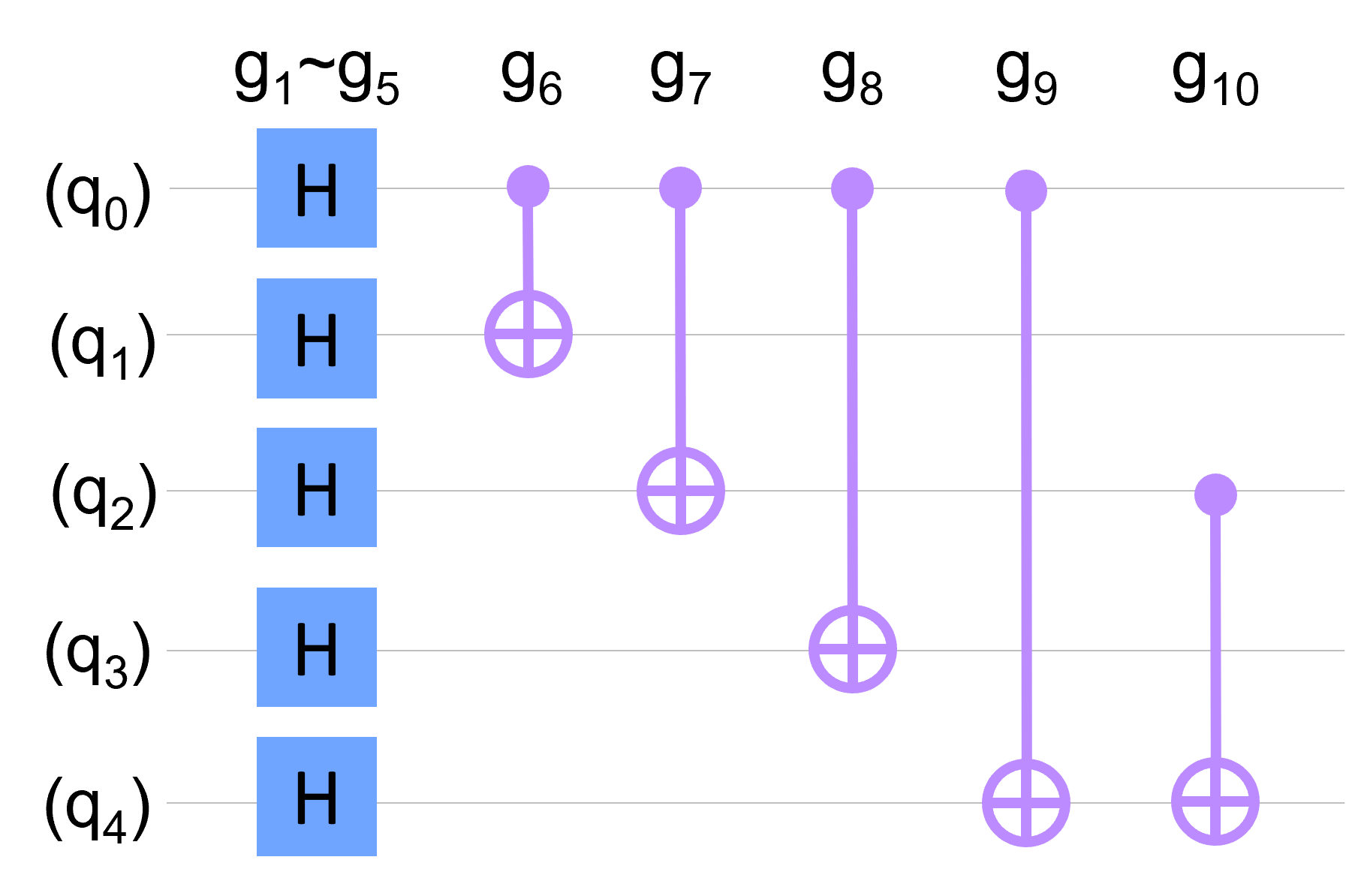} \label{qm_before}} 
		\hspace{0.01\columnwidth}
		\subfloat[The coupling graph of IBM QX2.]
		{\includegraphics[width=0.18\columnwidth]{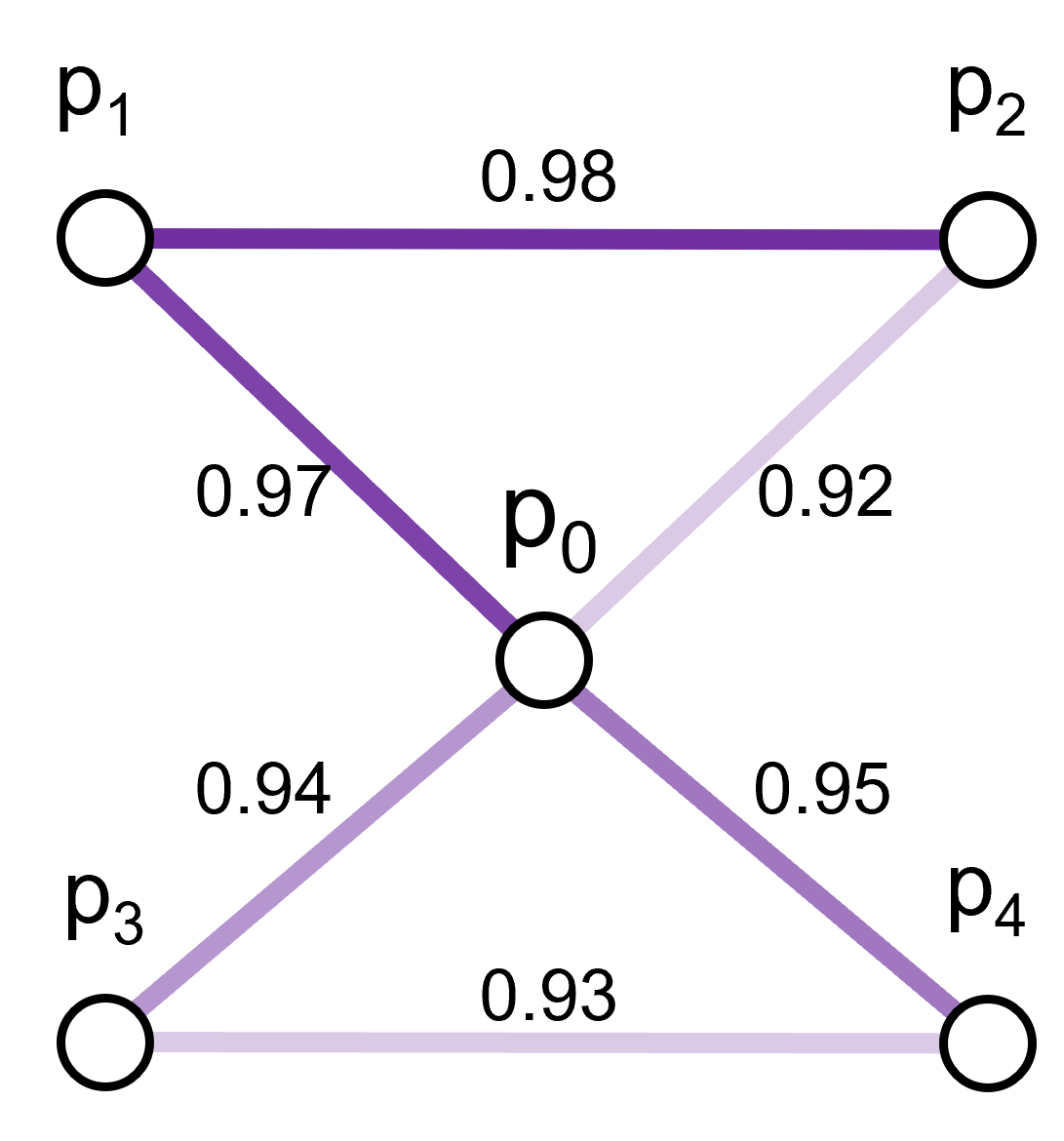} \label{qx}} 
		\hspace{0.01\columnwidth}
		\subfloat[Result of qubit mapping.]
		{\includegraphics[width=0.47\columnwidth]{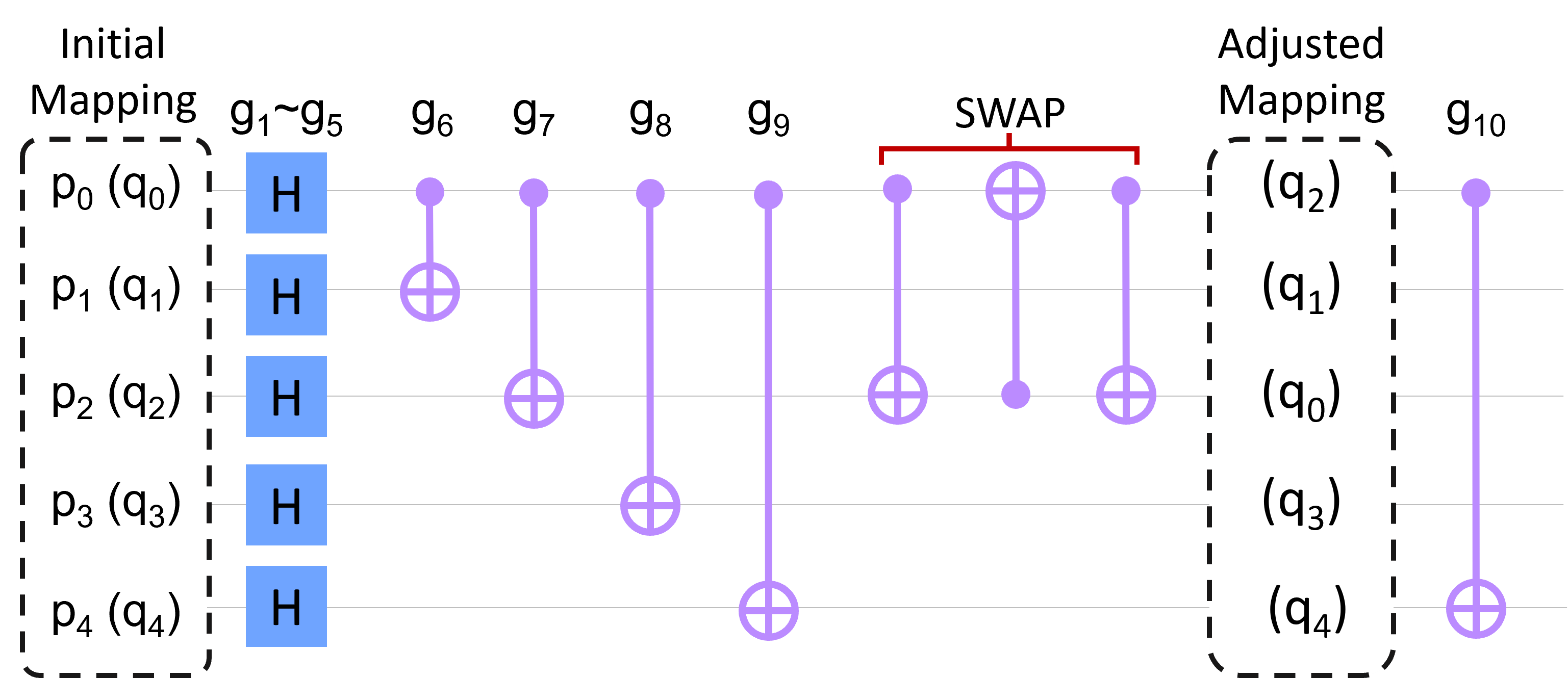} \label{qm_after}}
	\end{minipage}
	\caption{The depiction of qubit mapping.}
	\label{qm}
\end{figure}

Qubit mapping is the critical process of routing logical qubits from a quantum program to physical qubits within a quantum computer, ensuring that the mapped qubits satisfy the connectivity constraints of the target quantum hardware. Classical qubit mapping problems typically involve two primary inputs: the quantum program and the hardware-specific coupling graph. The coupling graph is formally represented as a graph $G = (P,E)$. $P$ denotes the set of physical qubits, $E$ defined as $(p_j,p_k)$, represents the set of connectivity edges , where $p_j$ and $p_k$ are distinct physical qubits connected within the quantum hardware's topology.
Key computational tasks of quantum qubit mapping workflow is as follows:

\begin{enumerate}
	\item{\textbf{Initial Mapping}\\
		Initial mapping represents the fundamental process of translating logical qubits from a quantum program to physical qubits on the connectivity graph at the circuit's inception. An optimal initial mapping can significantly enhance circuit fidelity, reduce circuit depth, and minimize swap operations. Formally, initial mapping can be represented as $m_{init}: q_0,q_1,q_2,q_3,q_4 \rightarrow p_0, p_1, p_2, p_3, p_4$.}\\
	
	\item{\textbf{Connectivity Compliance}\\
		Quantum circuit execution depends on the coupling graph's connectivity. When a quantum gate's required qubit connection isn't available, swap gates are used to rearrange qubits. SWAP gates allow dynamic repositioning of logical qubits to enable circuit execution. For instance, in Figure \ref{qm}, an initial mapping might prevent gate g$_{10}$ from running. By inserting a swap gate between physical qubits p$_0$ and p$_2$, the mapping can be adjusted to satisfy the connectivity requirements.}\\
	
	\item{\textbf{Fidelity}\\
		Fidelity serves as a critical metric in the NISQ era of quantum computing. At the gate level, two-qubit gates exhibit notably lower fidelity compared to single-qubit gates, representing a significant bottleneck in quantum circuit implementation \cite{Xu2020}. The fidelity of quantum gates can be determined through quantum process tomography \cite{Reich2013}. As illustrated in Figure \ref{qm}\subref{qx}, the fidelity of two-qubit gates is represented both numerically on the edges of the coupling graph and through the color depth of these edges, enabling researchers to effectively model noise effects. At the quantum circuit level, Hellinger Fidelity (HF) is commonly employed to measure the fidelity by quantifying the probabilistic deviation between noisy and ideal circuit outputs:
		
		\begin{equation}
			\label{hellinger}
			\text{HF}=(1-\frac{1}{2}\sum_{i=1}^{m}(\sqrt{o_p}-\sqrt{o_n})^2 )^{2}.
		\end{equation}
		Where $o_n$ and $o_p$ represent the output distributions of noisy and noise-free circuits respectively.
	}
\end{enumerate}

\subsection{Sovler-based qubit mapping}
Solvers, tools designed for solving Satisfiability Modulo Theories (SMT) or Boolean Satisfiability (SAT) problems, play a crucial role in qubit mapping tasks. In quantum circuit mapping, these solvers verify the existence of mapping strategies within given depth or swap constraints.
Key variables such as gate execution timing and mapping states are encoded as solver internal variables, with gate dependency relationships and mapping changes from swap gates represented as internal constraints. The solver iteratively adjusts circuit depth or swap requirements to validate solution existence, with the solution closest to satisfiability considered optimal. Qubit mapping has been proven to be an NP-complete problem\cite{Tan2021a}, leads to rapidly escalating solving complexity as circuit size increases. Experimental data from two state-of-the-art solver-based mapping methods, Qsynth-v2 and TB-OLSQ2, demonstrates this computational complexity.

As shown in Table\ref{tab:baseline-efficiency}, solving time increases dramatically with the number of two-qubit gates, rendering both methods unable to complete solving within a $\textbf{3600-second}$ time limit when two-qubit gate count exceeds $50$.

\begin{table}
	\small
	\renewcommand{\arraystretch}{0.95}
	\caption{Time consumption of two SOTA qubit mapping method.}
	\label{tab:baseline-efficiency}
	\centering
	\tabcolsep=0.02\linewidth
	\begin{tabular*}{\linewidth}{*{5}{c}}
		\toprule
		Samples & Qubits & 2Qu-gates & Qsynth-v2\cite{Shaik2024} & TB-OLSQ2\cite{Lin2023}\\
		\midrule
		qaoa5 &5 &8 & 0.706s &2.573s\\
		4mod5-v1\_22&5 &11 &2.166s &22.262s\\
		vqe\_8\_1\_5\_100 &6 &18 &5.303s &34.69s\\
		mod5mils\_65 &5 &16 &6.453s &93.755s\\
		barenco\_tof\_4 &7 &34 &40.48s &1380.9s\\
		vqe\_8\_4\_5\_100 &8 &39 &201.5s &1778.7s\\
		adder\_n10\_transpiled &10 &65 &2865.6s &\textbf{\textgreater 3600s}\\
		barenco\_tof\_5 &9 &50 &3111.02s &\textbf{\textgreater 3600s}\\
		vqe\_8\_2\_10\_100 &8 &79 &\textbf{\textgreater 3600s} &\textbf{\textgreater 3600s}\\
		vqe\_8\_3\_10\_100 &8 &78 &\textbf{\textgreater 3600s} &\textbf{\textgreater 3600s}\\
		\bottomrule
	\end{tabular*}
\end{table}

Furthermore, with the advancement of quantum circuit modeling techniques \cite{Huang2019}, researchers can now leverage measurement fidelities from real quantum computers to optimize quantum compilation \cite{Liu2020,Saravanan2023}. However, to our knowledge, existing solver-based qubit mapping methods have not yet effectively incorporated fidelity considerations into their approaches. This limitation largely stems from the inherent complexity of integrating continuous fidelity metrics into discrete solver formulations and the significant expansion of solution space that would result from considering fidelity variations across different hardware regions.

\section{The key issue of solver-based qubit mapping}

To further investigate the solver efficiency degradation issue, we conducted experiments to evaluate solver performance across quantum architectures of varying scales. As shown in Table \ref{tab:phy-qu-number-scalability}, the solving time of OLSQ2 solver\cite{Lin2023} demonstrates a substantial increase as the number of physical qubits grows in the quantum architecture.

\begin{table}
	\small
	\renewcommand{\arraystretch}{0.8}
	\caption{Solving time of OLSQ2 on different quantum computer.}
	\label{tab:phy-qu-number-scalability}
	\centering
	\tabcolsep=0.07\linewidth
	\begin{tabular*}{\linewidth}{*{4}{c}}
		\toprule
		Architecture & Aspen-4 & Grid 5x5 & Sycamore\\
		Qubits & 16 & 25 & 54\\
		\midrule
		simon\_n6 & 64s & 219s & 1119s\\
		basis\_test\_n4 &204s & 472s &1199s\\
		wstate\_n3 &30s &69s &271s\\
		fredkin\_n3&30s &88s &645s\\
		bv\_n14 & 188s & 152s & 605s\\
		ghz\_state\_n23 & N/A & 211s & 821s\\
		\bottomrule
	\end{tabular*}
\end{table}

This significant performance degradation can be attributed to two key factors:

\begin{enumerate}
	\item{\textbf{Variable Expansion}: The solver must evaluate the mapping possibilities of each logical qubit to every physical qubit at each timestep, leading to an expanded variable set as the number of physical qubits increases.}
	\item{\textbf{Constraint Proliferation}: The increase in physical qubits necessitates more constraints to ensure the correctness of the mapping, resulting in a proliferation of solver constraints.}
\end{enumerate}

These two factors jointly lead to an enlarged solution space and diminished solving efficiency. Furthermore, complexity analysis can be conducted to explore the root cause of solver efficiency degradation. To facilitate understanding of the subsequent complexity analysis, we summarize the key variables and their definitions in Table \ref{tab:variables}.

\begin{table}
	\small
	\renewcommand{\arraystretch}{0.95}
	\caption{Key variables and their definitions.}
	\label{tab:variables}
	\centering
	\tabcolsep=0.09\linewidth
	\begin{tabular*}{\linewidth}{*{2}{c}}
		\toprule
		Variables & Description\\
		\midrule
		$G$ & Set of two-qubit quantum gates\\
		$n_G$ & Number of two-qubit quantum gates\\
		$n_q$ & Number of logical qubits\\
		$n_{El}$ & Number of edges in the logical graph\\
		$B$ & Number of edges in the dependency chain\\
		$T$ & Upper bound of gate execution time\\
		$S$ & Upper bound of swap operations \\
		$P$ & Set of physical qubits in the coupling graph\\
		$n_P$ & Number of physical qubits in the coupling graph\\
		$E$ & Set of edges in the coupling graph\\
		$n_E$ & Number of edges in the coupling graph\\
		\bottomrule
	\end{tabular*}
\end{table}

These variables are used throughout our evaluation of the state-of-the-art solvers TB-OLSQ2 and Qsynth-v2, providing insights into the efficiency bottlenecks of solver-based qubit mapping approaches.

In the SAT problem domain, the number of variables, constraints, and clauses serve as crucial complexity indicators \cite{Sundermann2024}. According to \cite{Shaik2024}, the variable and clause numbers of Qsynth-v2 are:

\begin{equation}
	\label{n-var-qsynthv2}
	n_{var,qs-v2}=O(T(n_q \cdot n_P + n_{El} +n_G)).
\end{equation}

\begin{equation}
	\label{n-clause-qsynthv2}
	n_{clause,qs-v2}=O(T(n_q \cdot n_P + n_q \cdot n_E + n_{El}(n_P)^2 + n_G)).
\end{equation}

TB-OLSQ2 is a solver-based qubit mapping method based on the Coarse-Grained Circuit Model\cite{Lin2023}, comprising three distinct variable categories:

\begin{enumerate}
	\item $Mappings \ Variable (\pi_{q}^{t})$: Representing the logical qubit positions at each time step, with $T \times n_q$ variables.
	\item $Time \ Variables (t_g)$: Indicating the mapping time for each quantum gate, containing $G$ variables.
	\item $SWAP \ Variables (\sigma _{e}{t})$: Binary indicator variables for SWAP operations on graph edges at each time step, encompassing $T \times n_E$ variables.
\end{enumerate}

Finally, the total number of variables in the TB-OLSQ2 method can be expressed as:

\begin{equation}
	\label{n-var-tbolsq2}
	n_{var,tbolsq2}=O(T(n_q + n_E) + n_G).
\end{equation}

TB-OLSQ2 establishes comprehensive constraints to balance solving correctness and efficiency. These constraints include:

\begin{enumerate}
	\item $Injective \ Mapping \ Constraints$: These constraints prevent mapping conflicts among logical qubits and ensure the mapping scope. The number of constraints is $T(\frac{n_q(n_q-1)}{2}+2n_q)$ and can be simplified to $O(Tn_{q}^{2})$.
	\item $Consistency \ gate \ constraints$: These constraints guarantee that two-qubit gates are mapped onto physical edges in the connectivity graph. The number of constraints is $Tn_G(3n_E+1)$ and resulting in $O(Tn_Gn_E)$ for large $n_E$.
	\item $Dependency \ Constraints$: These constraints ensure the preservation of quantum gate execution order before and after mapping. The number of constraints is $B$.
	\item $SWAP \ Constraints$: These constraints ensure the non-overlapping nature of SWAP operations. The number of constraints can be expressed as $O(Nn_{E}d_{max}T)$.
	\item $Transformation \ Constraints$: These constraints facilitate the mapping and exchange functionality of SWAP operations. The number of constraints is $O(Tn_qn_P+Tn_qn_E)$.
\end{enumerate}

Consequently, the total number of constraints can be mathematically formulated as:

\begin{equation}
	\label{n-cons-tbolsq2}
	n_{cons,tbolsq2} = O(T(n_{q}^2 + (n_G+d_{max}+n_q)n_E + n_qn_P) + B).
\end{equation}

It is noteworthy that equations (\ref{n-var-qsynthv2})-(\ref{n-cons-tbolsq2}) reveal that variables $n_E$ and $n_P$, which are closely related to the hardware coupling graph scale, significantly affect the number of variables, clauses, and constraints in the solver. This theoretical analysis aligns with our experimental observations in Table \ref{tab:phy-qu-number-scalability}, where the solving time increases substantially with the growth of quantum architecture scale. The coupling graph expansion leads to considerable growth in both variable count and constraint number, resulting in significant increase in the solver's computational complexity. These findings confirm that \textbf{variable expansion} and \textbf{constraint proliferation} are indeed the key factors limiting solver efficiency, particularly for large-scale quantum architectures.

A potential optimization path emerges from this analysis: if the scale of the coupling graph involved in the solving process can be effectively reduced while maintaining mapping quality, the solver efficiency could be significantly improved. Based on this insight, we propose the HAQA method.

\section{HAQA Method} \label{haqa-method}
\subsection{HAQA Overview}

\begin{figure*}[htp]
	\centering
	\includegraphics[width=0.9\textwidth]{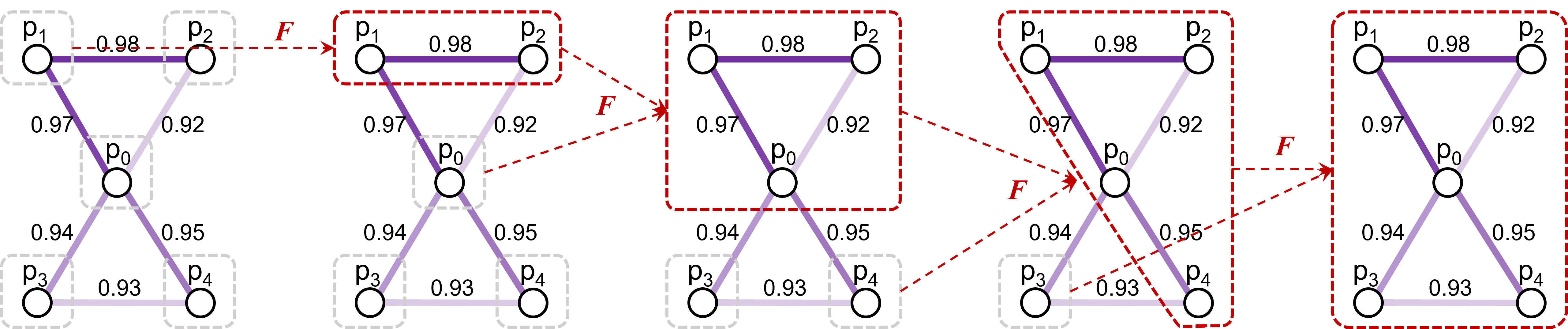}
	\caption{Recursive Community Fusion process on IBM QX2 coupling graph, each step illustrates community mergers (indicated by arrows) based on the reward function F. The process contains a record of the dominant community (highlighted in red) following each merger operation, with the process continuing iteratively until all nodes in the graph converge into a single unified community.}
	\label{recursive-ccommunity-fusion}
\end{figure*}

We observed that quantum circuits predominantly do not require the utilization of all physical qubits within the coupling graph during the mapping process. Let $n_q$ represent the number of logical qubits in the circuit, $n_P$ denote the total physical qubits in the coupling graph, and $n                                                                                                                                                                                                                                                                                             _m$ indicate the number of physical qubits involved in the mapping, with the constraint $n_q \leq n_m \leq n_P$. This observation provides a potential optimization space for the qubit mapping problem. Guo et al. \cite{Guo2024} similarly recognized this potential, employing a Subgraph Identification method to constrain the required physical qubits to the exact count of logical qubits, thereby achieving computational acceleration.
Extending this insight, HAQA emerges as a novel optimization method comprising two primary components:

\begin{enumerate}
	\item{\textbf{Recursive Community Fusion:} This component effectively integrates quantum circuit hardware information to construct a qubit-count-adaptive set of optimal regions on the coupling graph, aiming to generate mapping regions characterized by high connectivity.}
	\item{\textbf{Community Expansion:} This component subsequently enhances the algorithm's applicability through a strategic region expansion process, thereby broadening the method's potential for optimized quantum circuit mapping while maintaining solution quality.}
\end{enumerate}

\subsection{Recursive Community Fusion}

Recursive Community Fusion implements a community search methodology derived from the Fast Newman community detection algorithm \cite{Newman2004}. The algorithmic process, as depicted in Figure \ref{recursive-ccommunity-fusion}, initializes by establishing individual communities at each node of the coupling graph, where each node constitutes a distinct community. The algorithm then performs iterative community mergers guided by a reward function F that incorporates both connectivity and fidelity metrics. This iterative process continues until the coupling graph converges to a single unified community encompassing all physical qubits. For any two communities A and B, their merger potential is evaluated through the reward function defined as:

\begin{equation}
	\label{equ:f}
	F=Q+\omega E.
\end{equation}
Where $Q$ represents the modularity of a partition, defined as:
\begin{equation}
	\label{equ:Q}
	Q=\sum_{i}(p_{ii}-(\sum_{j}p_{ij})^2)
\end{equation}
$E$ represents the average two-qubit gate fidelity within the newly formed community:
\begin{equation}
	\label{equ:E}
	E = 1 - \frac{\sum_{l \in L_{in}} e_{l}}{\lvert{L_{in}}\rvert}
\end{equation}

Each execution generates a novel partition, enabling the computation of modularity $Q$ and average fidelity $E$. In modularity calculation, $p_{ii}$ denotes the probability of an edge residing within a community, while $p_{ij} (i\neq j)$ represents the probability of an edge connecting two distinct communities. A higher $Q$ value indicates stronger connectivity among nodes within the same community and weaker connections between nodes in different communities, signifying an improved partition.

The average fidelity $E$ calculation considers $l$ in $L_{in}$ as the edges connecting communities A and B, with $e_{l}$ representing two-qubit gate operational errors. To balance partition connectivity and fidelity, a weight parameter $w$ is introduced. When $w = 0$, the partitioning strategy solely considers connectivity, as $w$ increases, fidelity consideration becomes more prominent.

Throughout the fusion, the algorithm systematically evaluates each partition resulting from optimal merger strategies. The largest community in this partition is deemed the optimal community, and its corresponding subgraph in the coupling graph is preserved as a triple $(N_r, P_r, E_r)$, where $N_r$ represents the number of nodes in the community, $P_r$ denotes the complete set of nodes within the community, and $E_r$ comprises all edges interconnecting nodes within $P_r$ in the coupling graph. The above process is presented in Algorithm. \ref{alg:rcf}.

\begin{algorithm}
	\caption{Recursive Community Fusion}\label{alg:rcf}
	\begin{algorithmic}[1]
		\Require Coupling Graph $C = (P, E)$, where $P = \{p_1, p_2, ..., p_{n_P}\}$, $E = \{e_1, e_2, ..., e_{n_E}\}$, Edge Fidelity $K$
		\Ensure Triple Set $S_t = \emptyset$ 
		\State Max Reward $r = 0$
		\State Target Fusion Set $S = \emptyset$
		\State Community List $L_c = P$
			\While{$\lvert S\rvert \leq \lvert P\rvert$}
			\For{each pair $(set_i, set_j) \in L_c$, where $i \neq j$}
				\If{$F(L_c, set_i, set_j, K) \geq r$}
					\State $r$ = $F(L_c, set_i, set_j, K)$
					\State $S = set_i \cup set_j$
					\EndIf
				\EndFor
			\State $L_c$ = $L_c\setminus \{set_i,set_j\}$
			\State $L_c$.append($S$)
			\If{$S \notin \{a\vert (a,b,c) \in S_t\}$}
				\State $E_s=\{e\in E\vert$both endpoints of $e \in S\}$
				\State $S_t$.append($(\vert S\vert,S,E_s)$)
				\EndIf
			\EndWhile
		\State Return $S_t$
	\end{algorithmic}
\end{algorithm}

\subsection{Community Expansion} \label{community-expansion}

\begin{figure}[tb]
	\begin{minipage}[b]{\columnwidth}
		\centering
		\subfloat[The circuit needs mapping.]
		{\includegraphics[width=0.45\columnwidth]{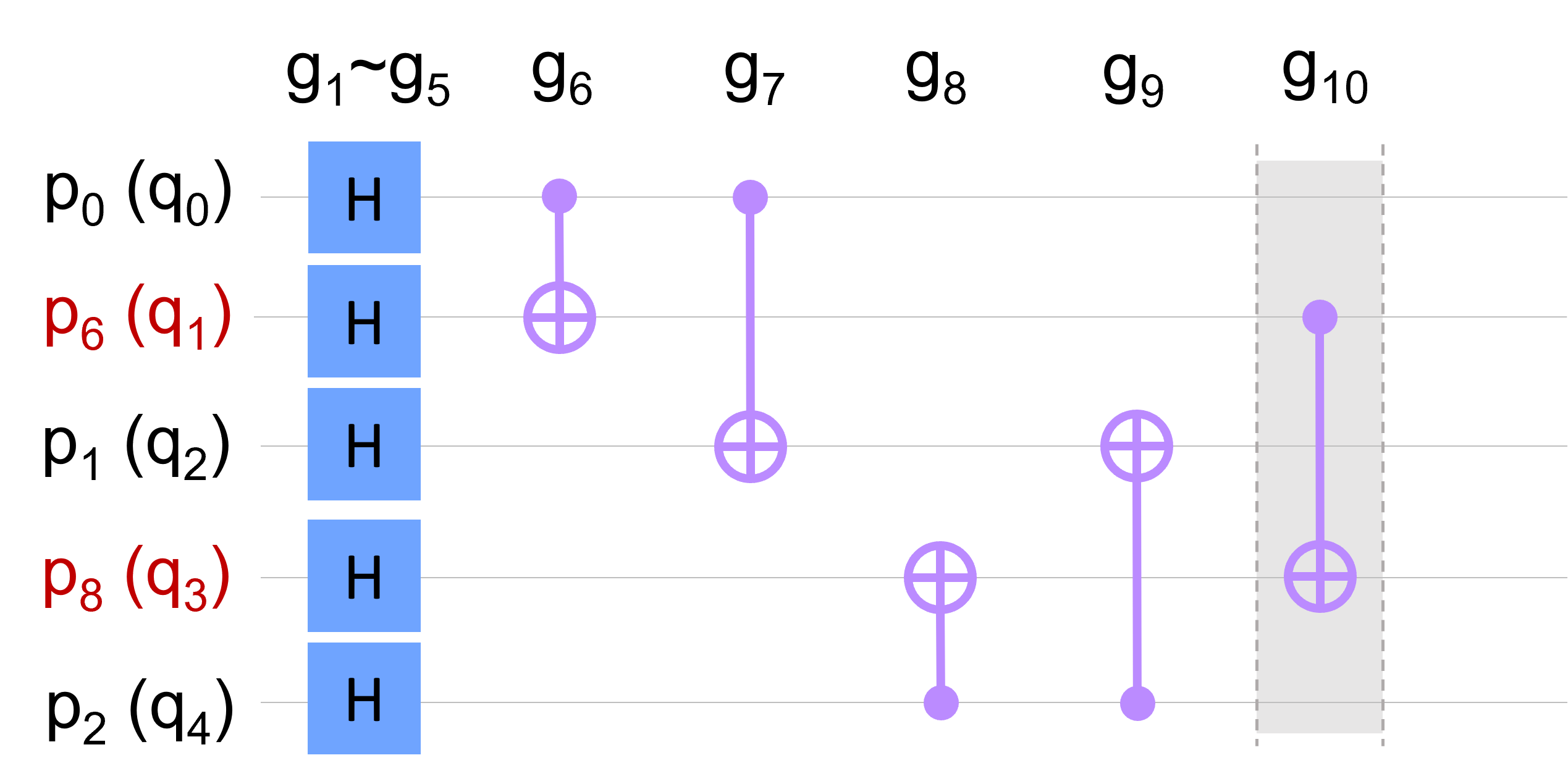} \label{com-expan-1-1}} 
		\hspace{0.01\columnwidth}
		\subfloat[SWAPs used in mapping.]
		{\includegraphics[width=0.45\columnwidth]{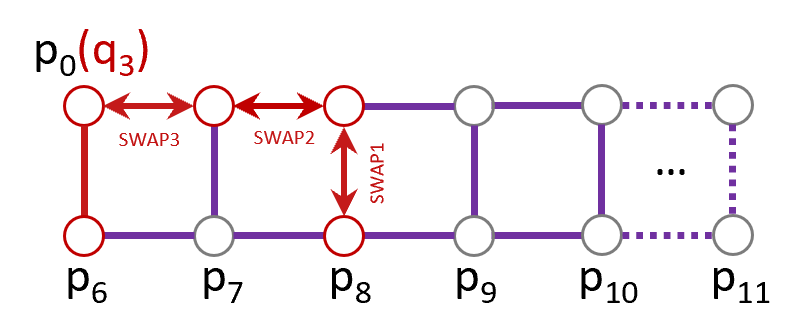} \label{com-expan-1-2}} 
	\end{minipage}
	\caption{Demonstration of additional swaps caused by a lack of sufficient auxiliary qubits.}
	\label{comm-expan-pre}
\end{figure}

The triple set obtained through the Recursive Community Fusion process encompasses regions with varying qubit counts. However, selecting regions with only the minimum required number of logical qubits frequently leads to insufficient auxiliary qubits for swap operations. This limitation not only severely impacts solving efficiency but often results in solving failures, as demonstrated in the following experimental analysis and illustrated in Figure \ref{comm-expan-pre}.

Consider the illustrative example, where quantum circuit is mapped onto a initial region with an equivalent number of qubits $p_0$, $p_6$, $p_1$, $p_8$, $p_2$. While the initial mapping $m_{init}: q_0,q_1,q_2,q_3,q_4$ $\rightarrow$ $p_0, p_6, p_1, p_8, p_2$ appears feasible for gates $g_1-g_9$, gate $g_{10}$ encounters connectivity constraints. Given the minimum distance between specific qubits being $4,$ at least $3$ swap gates are required for entanglement operations. These additional swap gates substantially increase the circuit complexity, not only diminishes circuit fidelity and increases circuit depth, but also leads to more search iterations in solver-based mapping methodologies, thereby compromising overall solving efficiency.

\begin{figure}
	\centering
	\includegraphics[width=0.5\textwidth]{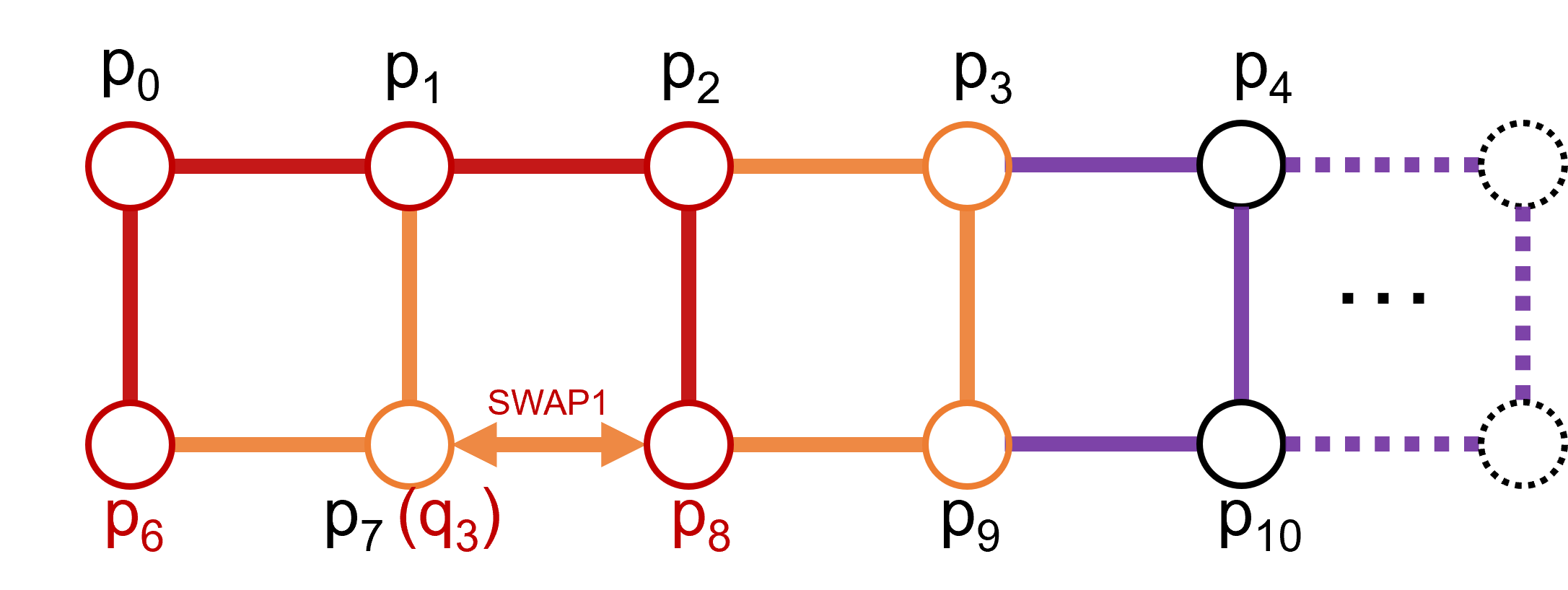}
	\caption{Demonstration of Community Expansion when $k=1$ (painted in orange).}
	\label{comm-expan}
\end{figure}

To address the challenge of insufficient auxiliary qubits, a straightforward solution is to incorporate additional physical qubits through community expansion, as illustrated in Fig. \ref{comm-expan}.

The expansion process enables convenient implementation of previously challenging swap operations by providing more flexible routing options.The expansion factor $k$ determines the iterative incorporation of adjacent physical qubits, where each expansion considers the quantum hardware's connectivity topology to include appropriate auxiliary qubits. With the integration of $p_7$ as an ancilla qubit during the $k=1$ expansion phase, the entire qubit mapping process requires only one additional swap operation to maintain connectivity requirements.

Community expansion is increasingly feasible with the rapid development of quantum computers providing more accessible physical qubits with good connectivity, and the hardware-aware region selection strategy ensures that the expanded regions maintain high fidelity characteristics throughout this process. The detailed algorithmic process is presented in Alg. \ref{alg:ce}.

\begin{algorithm}
	\caption{Community Expansion}
	\label{alg:ce}
	\begin{algorithmic}[1]
		\Require Triple Set $S_t$, Logical qubit number $n_q$, Physical qubit number $n_P$, Expansion rounds $k$
		\Ensure Final Mapping Graph $C_f=(P_f,E_f)$
		\State $P_f= \emptyset$
		\State $E_f= \emptyset$
		\State $N_r=n_P$
		\For{$T_r$ in $S_t$}
			\State $(a,b,c)=T_r$
			\If{$a\geq n_q$ and $a \leq N_r$}
			\State $N_r = a$
			\State $P_r = b$
			\State $E_r = c$
			\EndIf
			\EndFor
		\For{i = 1 to k}
			\For{$edge$ in $E_r$}
				\State Let $(p_x,p_y)$ be the two nodes connected by edge
				\If{$p_x$ in $S$ or $p_y$ in $S$}
					\State $P_f = P_f \cup \{P_x,P_y\}$
					\State $E_f = E_f \cup \{edge\}$
					\EndIf
				\EndFor
			\EndFor
		\State $C_f = (P_f,E_f)$
		\State Return $C_f$
	\end{algorithmic}
\end{algorithm}

\section{Complexity Analysis} \label{complexity-analysis}
Theoretical analysis demonstrates the computational complexity advantages of HAQA. Complexity analysis reveals polynomial-level acceleration of HAQA compared to TB-OLSQ2 \cite{Lin2023} and Qsynth-v2 \cite{Shaik2024}. All variables not specified in Table \ref{tab:variables} are detailed in Table \ref{tab:variables_new}.

\begin{table*}
	\scriptsize
	\renewcommand{\arraystretch}{1.2}
	\caption{Key variables and their definitions.}
	\label{tab:variables_new}
	\centering
	\tabcolsep=0.01\linewidth
	\begin{tabular*}{\linewidth}{@{\extracolsep{\fill}} c p{0.75\linewidth}}
		\toprule
		Variables & Description\\
		\midrule
		$C_f$ & Final Mapping Graph\\
		$n_P^b (n_P^w)$ & Number of physical qubits in the final mapping graph under optimal (worst) case\\
		$n_E^b (n_E^w)$ & Number of edges in the final mapping graph under optimal (worst) case\\
		$d_{max}(d_{min})$ & Maximum (Minimum) degree of the coupling graph\\
		$d_{max}^b (d_{max}^w)$ & Maximum degree of the coupling graph in the final mapping graph under optimal (worst) case\\
		$d_{min}^b (d_{min}^w)$ & Minimum degree of the coupling graph in the final mapping graph under optimal (worst) case\\
		$n_{var}^b (n_{var}^w)$ & Number of variables under optimal (worst) case after applying HAQA\\
		$n_{cons}^b (n_{cons}^w)$ & Number of constraints under optimal (worst) case after applying HAQA\\
		$n_{clause}^b (n_{clause}^w)$ & Number of clauses under optimal (worst) case after applying HAQA\\
		\bottomrule
	\end{tabular*}
\end{table*}

\begin{figure}[htbp]
	\centering
	\subfloat[The chain-like structure]{\includegraphics[width=0.35\columnwidth]{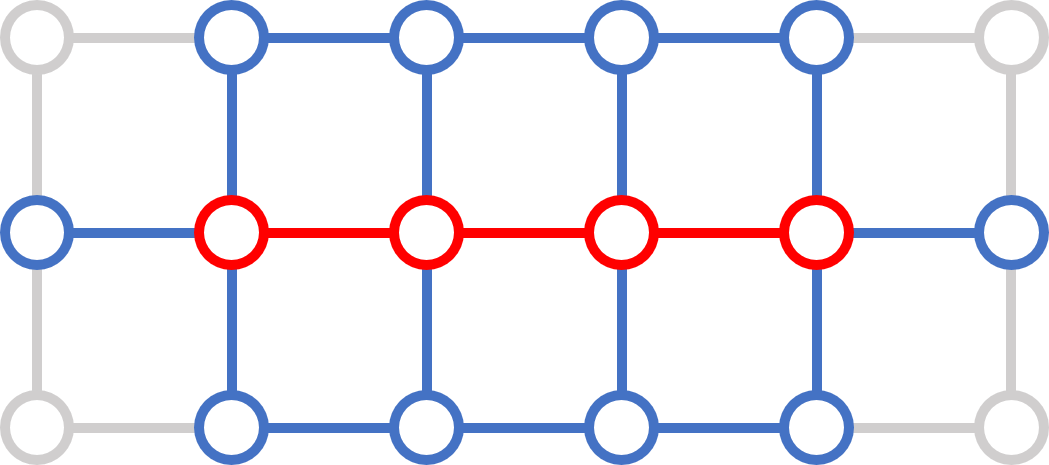}\label{fig:chain}}
	\hspace{0.05\linewidth}
	\subfloat[The pendent structure]{\includegraphics[width=0.27\columnwidth]{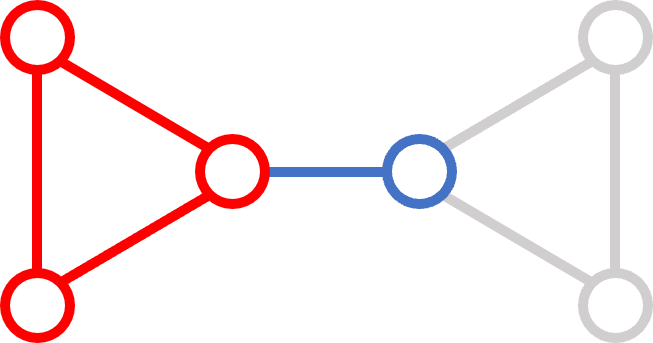} \label{fig:pendent}}
	\caption{Worst(chain-like) and optimal-case(pendent) scenarios, the red region represents the algorithm-determined qubits and edges, the blue region represents the neighboring qubits and adjacent edges expanded through community expansion at k=1, and the gray region represents other areas in the coupling graph.}
	\label{fig:two-structure}
\end{figure}

Most superconducting quantum computers adopt 2D structures, and the grid coupling graph being a commonly used and highly connected configuration. To avoid underestimating the complexity, we analyze the worst-case scenario on this structure, represented by a chain-like physical qubit region (Figure \ref{fig:two-structure}\subref{fig:chain}), which maximizes the number of qubits and physical edges. The red region represents the algorithm-determined qubits and edges, with $n_q$ qubits. The number of adjacent qubits does not exceed $n_q (d_{max}-1)+1$. The total number of physical qubits in the final mapped graph is equal to the sum of the quantities mentioned above: 

\begin{equation}
	\label{equ:np-worst}
	n_P^{w}\leq n_qd_{max}+1.
\end{equation}

Given $n_E^w$ and $d_{max}$, the upper bound of $n_E^{w}$ can be calculated as follows:

\begin{equation}
	\label{equ:ne-worst}
	n_E^{w}\leq \frac{d_{max}}{2}(n_qd_{max}+1).
\end{equation}

In the optimal scenario, the algorithm identifies a pendent subgraph of the entire coupling graph, characterized by minimal adjacent qubits and edges. A representative configuration is illustrated in Figure \ref{fig:two-structure}\subref{fig:pendent}. With the number of qubits in the final region $n_P^b$, we have $n_P^b = n_q + 1$. The edge number in the final region satisfies:

\begin{equation}
	\label{equ:ne-best}
	n_E^{b}\leq \frac{n_qd_{max}}{2}+1.
\end{equation}

Since the final coupling graph is a subset of the overall graph, we can assume the node degrees are similar:

\begin{equation}
	\label{equ:dmax}
	d_{max}^b = d_{max}^w = d_{max}
\end{equation}

\begin{equation}
	\label{equ:dmin}
	d_{min}^b = d_{min}^w = d_{min}
\end{equation}

Using the qubit numbers and edge counts derived from both the optimal and worst-case scenarios, we can compute the complexity of the improved method relative to different baselines. For the enhanced Qsynth-v2, the number of variables in the optimal and worst cases,  $n_{vars,qs-v2}^b$, $n_{vars,qs-v2}^w$ are:

\begin{equation}
	\label{equ:1}
	n_{vars,qs-v2}^b\leq O(T(n_q(n_q+1)+n_{El}+n_G))
\end{equation}

\begin{equation}
	\label{equ:2}
	n_{vars,qs-v2}^w\leq O(T(n_q(n_qd_{max}+1)+n_{El}+n_G))
\end{equation}

For the enhanced Qsynth-v2, the number of clauses in the optimal and worst cases,  $n_{clause,qs-v2}^b$, $n_{clause,qs-v2}^w$ are calculated as (\ref{equ:3}) and (\ref{equ:4}).

\begin{align}
	\label{equ:3}
	n_{clause,qs\text{-}v2}^b &\leq O\Big(T\Big(n_q(n_q + 1) + n_q\Big(\frac{n_q d_{max}}{2} + 1\Big) \notag \\
	&\quad + n_{El}(n_q + 1)^2 + n_G\Big)\Big) \\
	\label{equ:4}
	n_{clause,qs\text{-}v2}^w &\leq O\Big(T\Big(n_q(n_q d_{max} + 1) + n_q\Big(\frac{d_{max}}{2}(n_q d_{max} + 1)\Big) \notag \\
	&\quad + n_{El}(n_q d_{max} + 1)^2 + n_G\Big)\Big)
\end{align}

For the enhanced TB-OLSQ2, the number of variables in the optimal and worst cases,  $n_{vars,tbolsq2}^b$, $n_{vars,tbolsq2}^w$ are:

\begin{equation}
	\label{equ:tbolsq2-numvar-best}
	n_{var,tbolsq2}^b \leq T(\frac{n_qd_{max}}{2}+n_q+1)+n_G
\end{equation}

\begin{equation}
	\label{equ:tbolsq2-numvar-worst}
	n_{var,tbolsq2}^b \leq T(\frac{d_{max}}{2}(n_qd_{max}+1)+n_q)+n_G
\end{equation}

For the enhanced TB-OLSQ2, the number of constraints in the optimal and worst cases,  $n_{cons,tbolsq2}^b$, $n_{cons,tbolsq2}^w$ are calculated as (\ref{equ:tbolsq2-numcons-best}) and (\ref{equ:tbolsq2-numcons-worst}).

\begin{align}
	\label{equ:tbolsq2-numcons-best}
	n_{cons,tbolsq2}^b &\leq O\Big(T\Big(n_q^2 + (n_G + d_{max} + n_q)\Big(\frac{n_q d_{max}}{2} + 1\Big) \notag \\
	&\quad + n_q(n_q + 1)\Big) + B\Big) \\
	\label{equ:tbolsq2-numcons-worst}
	n_{cons,tbolsq2}^w &\leq O\Big(T\Big(n_q^2 + (n_G + d_{max} + n_q)\Big(\frac{d_{max}(n_q d_{max} + 1)}{2}\Big) \notag \\
	&\quad + n_q(n_q d_{max} + 1)\Big) + B\Big)
\end{align}

We define the predicted average number of physical qubits after improvement, denoted as $n_P^{avg}$, as the mean of its upper and lower bounds. Similarly, the predicted average number of edges after improvement, denoted as $n_E^{avg}$, is defined as the mean of its corresponding bounds:

\begin{equation}
	\label{equ:nq-avg}
	n_P^{avg} = \frac{n_P^b + n_P^w}{2}\leq \frac{n_q(d_{max}+1)}{2}+1
\end{equation}

\begin{equation}
	\label{equ:ne-avg}
	n_E^{avg} = \frac{n_E^b + n_E^w}{2}
\end{equation}

Based on $n_P^{avg}$ and $n_E^{avg}$, we introduce two performance indicators:

\begin{enumerate}
	\item{\textbf{Average Qubit Pruning Ratio ($r_{aq}$):}
		\begin{equation}
			\label{equ:aqpr}
			r_{aq}=\frac{n_P}{n_P^{avg}}
		\end{equation}
	}
	\item{\textbf{Average Edge Pruning Ratio ($r_{ae}$):}
		\begin{equation}
			\label{equ:aepr}
			r_{ae}=\frac{n_E}{n_E^{avg}}
		\end{equation}
	}
\end{enumerate}

Using $r_{aq}$ and $r_{ae}$, the variable and clause counts for Qsynth-v2 after applying HAQA can be expressed as (\ref{equ:haqa-var-qsv2}) and (\ref{equ:haqa-clause-qsv2}), the variable and clause counts for TB-OLSQ2 after applying HAQA can be expressed as (\ref{equ:haqa-var-tbolsq2}) and (\ref{equ:haqa-cons-tbolsq2}).

\begin{equation}
	\begin{aligned}
		\label{equ:haqa-var-qsv2}
		n_{var,qs\text{-}v2}^{\text{HAQA}} &= O\Big(T\Big(\frac{n_q n_P}{r_{aq}} + n_{El} + n_G\Big)\Big)
	\end{aligned}
\end{equation}

\begin{equation}
	\begin{aligned}
		\label{equ:haqa-clause-qsv2}
		n_{clause,qs\text{-}v2}^{\text{HAQA}} = O\Big(T\Big(\frac{n_q n_P}{r_{aq}} + \frac{n_q n_E}{r_{ae}} \quad + \frac{n_{El} (n_P)^2}{r_{aq}^2} + n_G\Big)\Big)
	\end{aligned}
\end{equation}

\begin{equation}
	\begin{aligned}
		\label{equ:haqa-var-tbolsq2}
		n_{var,tbolsq2}^{\text{HAQA}} = T\Big(n_q + \frac{n_E}{r_{ae}}\Big) + n_G
	\end{aligned}
\end{equation}

\begin{equation}
	\begin{aligned}
		\label{equ:haqa-cons-tbolsq2}
		n_{cons,tbolsq2}^{\text{HAQA}} = O\Big(T\Big(n_q^2 + \frac{(n_G + d_{max} + n_q) n_E}{r_{ae}} \quad + \frac{n_q n_P}{r_{aq}}\Big) + B\Big)
	\end{aligned}
\end{equation}

Fundamentally, HAQA reduces the coupling graph-related parameters $n_E$ and $n_P$ to the order of logical qubit count $n_q$ in quantum circuits, thereby significantly improving the solution efficiency. Based on (\ref{equ:haqa-var-qsv2})-(\ref{equ:haqa-cons-tbolsq2}), $r_{aq}$ and $r_{ae}$ enable a comprehensive assessment of HAQA's impact on variable, clause, and constraint complexity. For Qsynth-v2, HAQA reduces the first term of variable complexity to a first-order polynomial level. Regarding clause complexity, the algorithm achieves a first-order polynomial reduction in the first and second term and a second-order polynomial reduction in the third term of the complexity formula. Similarly, for TB-OLSQ2, HAQA reduces the second term of variable complexity to a first-order polynomial level, and reduces the second and third terms of constraint complexity to a first-order polynomial level. The complexity analysis methodology presented here offers a generalizable framework for evaluating solver-based qubit mapping methods, applicable to various quantum architectures and mapping strategies. This analytical approach provides quantitative insights into the performance gains of optimization techniques, establishing a foundation for objective assessment and continuous improvement of qubit mapping methodologies in quantum compilation workflows.

Through theoretical analysis, HAQA demonstrates its effectiveness in addressing the key challenges in solver-based qubit mapping. The method establishes hardware-aware mapping regions through \textbf{Recursive Community Fusion}, with \textbf{Community Expansion} ensuring solution quality by adaptively incorporating auxiliary qubits. This approach reduces computational complexity to a polynomial level by controlling both variable expansion and constraint proliferation, while enabling quantum fidelity optimization through hardware-characteristic-based region selection. The subsequent numerical experiments provide further validation of the method's practical performance improvements.

\section{Experimental Investigation and Analysis}
\subsection{Experimental Setup}
\textbf{Metrics}:
We evaluate the efficiency and mapping performance of HAQA using the following metrics. First, solving efficiency is assessed by recording and comparing the runtime of various circuits across different qubit mapping methods. A time limit of 3600 seconds (1 hour) is set for each sample, with any method exceeding this limit considered a failure to solve, denoted as TO in experiment results.
Hellinger Fidelity is used as the primary performance indicator for the post-mapping circuit. Using Qiskit\cite{qiskit}, we simulate two scenarios:
\begin{enumerate}
	\item Noisy Circuit: The circuit after qubit mapping, simulated in an environment with two-qubit gate errors, using noise data provided by IBM.
	\item Ideal Circuit: The original circuit (prior to qubit mapping), simulated in a noise-free environment.
\end{enumerate}
In both scenarios, the circuits are executed 1024 times, and the resulting distributions are used to compute the Hellinger Fidelity.
Circuit depth and SWAP count are critical evaluation metrics for solver-based qubit mapping methods. Comparing these indicators allows us to determine whether HAQA retains the inherent advantages of the baseline mapping approaches.

\textbf{Baselines}:
Our baselines are two state-of-the-art solver-based qubit mapping methods: QSynth-v2 and TB-OLSQ2. QSynth-v2 focuses on parallel plans and domain-specific information for mapping circuits to platforms with over 100 qubits, while TB-OLSQ2 emphasizes scalability through concise SMT formulations and iterative optimizations.

\textbf{Benchmark Circuits}:
We utilize 25 quantum circuits from QSynth-v2\cite{Shaik2024} and TB-OLSQ2\cite{Lin2023}, with circuit depths ranging from 8 to 136. Due to memory constraints, we focus on quantum circuits with 3 to 10 qubits.

\textbf{Quantum Computer Setup}:
To evaluate the solution efficiency and quality of our method, we use hardware information (coupling graph and fidelity data) from IBM Eagle and IBM Heron for testing. IBM Eagle consists of 127 qubits, while IBM Heron has 133 qubits, both featuring a heavy hex lattice coupling graph. In terms of fidelity, IBM Heron has a slightly lower two-qubit gate error compared to IBM Eagle. The coupling graphs of IBM Eagle and IBM Heron is shown in Appendix \ref{apdix}.

\textbf{Hardware/Software Setup}:
All of our experiments were run on a Ryzen7 5700G CPU running at 3.8 GHz with 32 GB of RAM. HAQA was implemented using Networkx (v3.4.2). Due to different environment requirements for the two baselines, the experiments for TB-OLSQ2 and its application of the HAQA algorithm were conducted in Python 3.9, using the Z3 solver (v4.12.5.0) and Pysat (v0.1.8.dev12). The experiments for QSynth v2 and its application of HAQA were conducted in Python 3.10, using Qiskit (v0.46.3) and Pysat (v1.8.dev13).

\textbf{Expansion Factor Determination}:
As mentioned in subsection \ref{community-expansion}, the selection of expansion factor k directly impacts both the efficiency and success rate of the mapping process. A series of preliminary experiments were conducted to determine the optimal k value, with results presented in Table \ref{tab:determine-k}.

\begin{table}[htbp]
	\small
	\renewcommand{\arraystretch}{1}
	\caption{Solving Time of HAQA-TB-OLSQ2 on IBM Heron for Varying k.}
	\label{tab:determine-k}
	\centering
	\tabcolsep=0.035\linewidth
	\begin{tabular*}{\linewidth}{*{5}{c}}
		\toprule
		Samples            & TB-OLSQ2 & HAQA$_{k=0}$ & HAQA$_{k=1}$ & HAQA$_{k=2}$ \\
		\midrule
		vqe\_8\_0\_10\_100 & TO       & TO           & \textbf{468.04}       & 1850.78      \\
		vqe\_8\_0\_5\_100  & TO       & TO           & \textbf{112.7}        & 304.11       \\
		vqe\_8\_1\_10\_100 & TO       & TO           & \textbf{134.49}       & 207.21       \\
		vqe\_8\_2\_10\_100 & TO       & TO           & \textbf{1330.59}      & TO           \\
		vqe\_8\_2\_5\_100  & TO       & 1133.72      & \textbf{56.65}        & 84.83        \\
		vqe\_8\_3\_10\_100 & TO       & TO           & TO           & TO           \\
		vqe\_8\_3\_5\_100  & TO       & 1878.67      & \textbf{50.53}        & 104.22       \\
		vqe\_8\_4\_10\_100 & TO       & TO           & TO           & TO           \\
		\bottomrule
	\end{tabular*}
\end{table}

The experimental data demonstrates that $k=1$ achieves optimal performance, exhibiting minimal circuit solving time and the highest number of successfully solved circuits within the time constraint. Performance degradation at $k=2$ can be attributed to the increased computational complexity from larger initial regions containing more physical qubits. At $k=0$, configurations demonstrate both significantly extended solving times and frequent solving failures, likely due to insufficient auxiliary qubits necessitating additional swap operations, thereby increasing circuit depth and complicating the search process.

\subsection{Efficiency}

Tables \ref{tab:time-ibmeagle} and \ref{tab:time-ibmheron} present the solving time and acceleration ratios for QSynth-v2 and TB-OLSQ2 with HAQA implementation on IBM Eagle and IBM Heron platforms. The acceleration ratio (Acc-Ratio) is calculated using the time limit of 3600s for cases exceeding the solving time limit. HAQA demonstrates acceleration across all test cases.\\
On IBM Eagle, HAQA-enhanced QSynth-v2 achieves a maximum speedup of 632.76x and an average speedup greater than 182.9x, while HAQA-enhanced TB-OLSQ2 shows a maximum speedup of 197.06x and an average speedup exceeding 57.54x. Furthermore, with HAQA acceleration, both QSynth-v2 and TB-OLSQ2 complete all test cases within the time limit, indicating substantial improvement in solving efficiency.
On IBM Heron, HAQA enables QSynth-v2 to reach a maximum speedup of 580.36x with an average speedup greater than 149.95x, while TB-OLSQ2 achieves a maximum speedup of 286.67x and an average speedup exceeding 61.21x. HAQA-enhanced QSynth-v2 completes all test cases within the time limit, while HAQA-enhanced TB-OLSQ2 completes the majority of test cases within the specified time constraint.
Notably, on both quantum computers, QSynth-v2 shows higher average speedup than TB-OLSQ2, which aligns with our complexity analysis in Section \ref{complexity-analysis}, where HAQA achieves quadratic polynomial acceleration for QSynth-v2 and linear polynomial acceleration for TB-OLSQ2.

\begin{table*}[htbp]
	\small
	\caption{Solving Time on IBM Eagle.}
	\label{tab:time-ibmeagle}
	\centering
	\tabcolsep=0.01\linewidth
	\renewcommand{\arraystretch}{1.1}
	\resizebox{\textwidth}{!}{
	\begin{tabular}{ccccccrccr}
		\toprule
		\multirow{2}{*}{Samples} & \multirow{2}{*}{Depth} & \multirow{2}{*}{$n_q$} & \multirow{2}{*}{$n_G$} & \multicolumn{3}{c}{QSynth-v2 (s)} & \multicolumn{3}{c}{TB-OLSQ2 (s)} \\
		\cmidrule(lr){5-7}\cmidrule(lr){8-10}
		& & & & Baseline & HAQA(ours) & \textbf{Acc-Ratio} & Baseline  & HAQA(ours) & \textbf{Acc-Ratio}   \\
		\midrule
		4gt13\_92              & 38  & 5  & 30 & 357.11  & 0.702  & \textbf{508.87$\enspace\uparrow$}               & TO      & 21.35  & \textbf{\textgreater{}168.65$\enspace\uparrow$} \\
		4mod5-v1\_22           & 12  & 5  & 11 & 2.17    & 0.024  & \textbf{89.41$\enspace\uparrow$}                & 22.26   & 0.352  & \textbf{63.25$\enspace\uparrow$}                \\
		adder                  & 11  & 4  & 10 & 2.98    & 0.019  & \textbf{155.08$\enspace\uparrow$}               & 5.11    & 0.161  & \textbf{31.68$\enspace\uparrow$}                \\
		adder\_n10\_transpiled & 119 & 10 & 65 & 2865.64 & 11.37  & \textbf{251.96$\enspace\uparrow$}               & TO      & 478.01 & \textbf{\textgreater{}7.53$\enspace\uparrow$}   \\
		barenco\_tof\_4        & 68  & 7  & 34 & 40.49   & 0.529  & \textbf{76.57$\enspace\uparrow$}                & 1380.94 & 17.79  & \textbf{77.63$\enspace\uparrow$}                \\
		barenco\_tof\_5        & 95  & 9  & 50 & 3111.02 & 4.92   & \textbf{632.76$\enspace\uparrow$}               & TO      & 57.81  & \textbf{\textgreater{}62.27$\enspace\uparrow$}  \\
		mod\_mult\_55          & 47  & 9  & 40 & TO      & 10.41  & \textbf{\textgreater{}345.79$\enspace\uparrow$} & TO      & 103.95 & \textbf{\textgreater{}34.63$\enspace\uparrow$}  \\
		mod5mils\_65           & 21  & 5  & 16 & 6.45    & 0.05   & \textbf{130.25$\enspace\uparrow$}               & 93.75   & 2.88   & \textbf{32.51$\enspace\uparrow$}                \\
		or                     & 8   & 3  & 6  & 1.33    & 0.011  & \textbf{118.2$\enspace\uparrow$}                & 7.7     & 0.109  & \textbf{70.95$\enspace\uparrow$}                \\
		qaoa5                  & 14  & 5  & 8  & 0.706   & 0.01   & \textbf{69.07$\enspace\uparrow$}                & 2.57    & 0.189  & \textbf{13.61$\enspace\uparrow$}                \\
		qft\_8                 & 42  & 8  & 56 & TO      & 1296.2 & \textbf{\textgreater{}2.78$\enspace\uparrow$}   & TO      & 110.82 & \textbf{\textgreater{}32.49$\enspace\uparrow$}  \\
		qpe\_n9\_transpiled    & 92  & 9  & 43 & 1269.92 & 3.16   & \textbf{402.32$\enspace\uparrow$}               & TO      & 55.16  & \textbf{\textgreater{}65.26$\enspace\uparrow$}  \\
		tof\_4                 & 46  & 7  & 22 & 6.57    & 0.039  & \textbf{167.29$\enspace\uparrow$}               & 47.99   & 0.661  & \textbf{72.55$\enspace\uparrow$}                \\
		tof\_5                 & 61  & 9  & 30 & 37.7    & 0.38   & \textbf{99.31$\enspace\uparrow$}                & 2299.11 & 11.67  & \textbf{197.06$\enspace\uparrow$}               \\
		vbe\_adder\_3          & 58  & 10 & 50 & 644.45  & 5.56   & \textbf{116.01$\enspace\uparrow$}               & TO      & 35.71  & \textbf{\textgreater{}100.8$\enspace\uparrow$}  \\
		vqe\_8\_0\_10\_100     & 92  & 8  & 63 & TO      & 72.39  & \textbf{\textgreater{}49.73$\enspace\uparrow$}  & TO      & 181.15 & \textbf{\textgreater{}19.87$\enspace\uparrow$}  \\
		vqe\_8\_0\_5\_100      & 79  & 8  & 52 & TO      & 113.15 & \textbf{\textgreater{}31.82$\enspace\uparrow$}  & TO      & 93.9   & \textbf{\textgreater{}38.34$\enspace\uparrow$}  \\
		vqe\_8\_1\_10\_100     & 76  & 7  & 47 & TO      & 6.59   & \textbf{\textgreater{}546.17$\enspace\uparrow$} & TO      & 63.13  & \textbf{\textgreater{}57.03$\enspace\uparrow$}  \\
		vqe\_8\_1\_5\_100      & 32  & 6  & 18 & 5.3     & 0.044  & \textbf{119.28$\enspace\uparrow$}               & 34.69   & 0.606  & \textbf{57.25$\enspace\uparrow$}                \\
		vqe\_8\_2\_10\_100     & 136 & 8  & 79 & TO      & 41.02  & \textbf{\textgreater{}87.76$\enspace\uparrow$}  & TO      & 488.21 & \textbf{\textgreater{}7.37$\enspace\uparrow$}   \\
		vqe\_8\_2\_5\_100      & 80  & 8  & 48 & 1442.87 & 4.05   & \textbf{356.09$\enspace\uparrow$}               & TO      & 41.73  & \textbf{\textgreater{}86.28$\enspace\uparrow$}  \\
		vqe\_8\_3\_10\_100     & 119 & 8  & 78 & TO      & 92.82  & \textbf{\textgreater{}38.79$\enspace\uparrow$}  & TO      & 1018.7 & \textbf{\textgreater{}3.53$\enspace\uparrow$}   \\
		vqe\_8\_3\_5\_100      & 61  & 8  & 40 & 467.66  & 5.95   & \textbf{78.59$\enspace\uparrow$}                & TO      & 43.63  & \textbf{\textgreater{}82.52$\enspace\uparrow$}  \\
		vqe\_8\_4\_10\_100     & 102 & 8  & 71 & TO      & 1333.1 & \textbf{\textgreater{}2.7$\enspace\uparrow$}    & TO      & 257.5  & \textbf{\textgreater{}13.98$\enspace\uparrow$}  \\
		vqe\_8\_4\_5\_100      & 60  & 8  & 39 & 201.56  & 2.1    & \textbf{95.81$\enspace\uparrow$}                & 1778.66 & 42.98  & \textbf{41.38$\enspace\uparrow$}                \\
		Average                &-&-&-&-&-& \textbf{\textgreater{}182.9$\enspace\uparrow$}  &-&-& \textbf{\textgreater{}57.54$\enspace\uparrow$}   \\
		\bottomrule
	\end{tabular}
	}
\end{table*}

\begin{table*}[htbp]
	\small
	\caption{Solving Time on IBM Heron.}
	\label{tab:time-ibmheron}
	\centering
	\tabcolsep=0.015\linewidth
	\renewcommand{\arraystretch}{1.1}
	\resizebox{\textwidth}{!}{
	\begin{tabular}{ccccccrccr}
		\toprule
		\multirow{2}{*}{Samples} & \multirow{2}{*}{Depth} & \multirow{2}{*}{$n_q$} & \multirow{2}{*}{$n_G$} & \multicolumn{3}{c}{QSynth-v2 (s)} & \multicolumn{3}{c}{TB-OLSQ2 (s)} \\
		\cmidrule(lr){5-7}\cmidrule(lr){8-10}
		& & & & Baseline & HAQA(ours) & \textbf{Acc-Ratio} & Baseline  & HAQA(ours) & \textbf{Acc-Ratio}   \\
		\midrule
		4gt13\_92              & 38  & 5  & 30 & 419.82  & 0.723   & \textbf{580.36$\enspace\uparrow$}               & TO      & 17.26   & \textbf{\textgreater{}208.52$\enspace\uparrow$} \\
		4mod5-v1\_22           & 12  & 5  & 11 & 3.49    & 0.02    & \textbf{171.63$\enspace\uparrow$}               & 127.7   & 0.445   & \textbf{286.87$\enspace\uparrow$}               \\
		adder                  & 11  & 4  & 10 & 2.96    & 0.019   & \textbf{155.27$\enspace\uparrow$}               & 10.83   & 0.32    & \textbf{33.85$\enspace\uparrow$}                \\
		adder\_n10\_transpiled & 119 & 10 & 65 & 3452.2  & 17.44   & \textbf{197.97$\enspace\uparrow$}               & TO      & TO      & N/A -                                    \\
		barenco\_tof\_4        & 68  & 7  & 34 & 53.05   & 0.721   & \textbf{73.62$\enspace\uparrow$}                & 2341.18 & 25.23   & \textbf{92.8$\enspace\uparrow$}                 \\
		barenco\_tof\_5        & 95  & 9  & 50 & 2967.75 & 14.92   & \textbf{198.92$\enspace\uparrow$}               & TO      & 226.33  & \textbf{\textgreater{}15.91$\enspace\uparrow$}  \\
		mod\_mult\_55          & 47  & 9  & 40 & TO      & 70.06   & \textbf{\textgreater{}51.38$\enspace\uparrow$}  & TO      & 244.65  & \textbf{\textgreater{}14.72$\enspace\uparrow$}  \\
		mod5mils\_65           & 21  & 5  & 16 & 5.95    & 0.04    & \textbf{146.98$\enspace\uparrow$}               & 188.16  & 3.52    & \textbf{53.42$\enspace\uparrow$}                \\
		or                     & 8   & 3  & 6  & 2.15    & 0.012   & \textbf{174.99$\enspace\uparrow$}               & 11.7    & 0.213   & \textbf{54.91$\enspace\uparrow$}                \\
		qaoa5                  & 14  & 5  & 8  & 0.785   & 0.009   & \textbf{84.12$\enspace\uparrow$}                & 5.86    & 0.272   & \textbf{21.51$\enspace\uparrow$}                \\
		qft\_8                 & 42  & 8  & 56 & TO      & 487.48  & \textbf{\textgreater{}7.38$\enspace\uparrow$}   & TO      & 174.48  & \textbf{\textgreater{}20.63$\enspace\uparrow$}  \\
		qpe\_n9\_transpiled    & 92  & 9  & 43 & 1368.61 & 8.89    & \textbf{153.94$\enspace\uparrow$}               & TO      & 210.45  & \textbf{\textgreater{}17.11$\enspace\uparrow$}  \\
		tof\_4                 & 46  & 7  & 22 & 10.02   & 0.054   & \textbf{185.16$\enspace\uparrow$}               & 70.52   & 1.31    & \textbf{53.78$\enspace\uparrow$}                \\
		tof\_5                 & 61  & 9  & 30 & 55.76   & 0.75    & \textbf{74.3$\enspace\uparrow$}                 & TO      & 40.54   & \textbf{\textgreater{}88.79$\enspace\uparrow$}  \\
		vbe\_adder\_3          & 58  & 10 & 50 & 520.93  & 4.45    & \textbf{116.99$\enspace\uparrow$}               & TO      & 78.54   & \textbf{\textgreater{}45.84$\enspace\uparrow$}  \\
		vqe\_8\_0\_10\_100     & 92  & 8  & 63 & TO      & 55.37   & \textbf{\textgreater{}65.02$\enspace\uparrow$}  & TO      & 468.04  & \textbf{\textgreater{}7.69$\enspace\uparrow$}   \\
		vqe\_8\_0\_5\_100      & 79  & 8  & 52 & TO      & 109.93  & \textbf{\textgreater{}32.75$\enspace\uparrow$}  & TO      & 112.7   & \textbf{\textgreater{}31.94$\enspace\uparrow$}  \\
		vqe\_8\_1\_10\_100     & 76  & 7  & 47 & TO      & 12.84   & \textbf{\textgreater{}280.34$\enspace\uparrow$} & TO      & 134.49  & \textbf{\textgreater{}26.77$\enspace\uparrow$}  \\
		vqe\_8\_1\_5\_100      & 32  & 6  & 18 & 6.47    & 0.056   & \textbf{115.21$\enspace\uparrow$}               & 33.3    & 1.38    & \textbf{24.16$\enspace\uparrow$}                \\
		vqe\_8\_2\_10\_100     & 136 & 8  & 79 & TO      & 193.64  & \textbf{\textgreater{}18.59$\enspace\uparrow$}  & TO      & 1330.59 & \textbf{\textgreater{}2.71$\enspace\uparrow$}   \\
		vqe\_8\_2\_5\_100      & 80  & 8  & 48 & 2478.65 & 5.82    & \textbf{425.87$\enspace\uparrow$}               & TO      & 56.65   & \textbf{\textgreater{}63.54$\enspace\uparrow$}  \\
		vqe\_8\_3\_10\_100     & 119 & 8  & 78 & TO      & 1514.45 & \textbf{\textgreater{}2.38$\enspace\uparrow$}   & TO      & TO      & N/A -                  \\
		vqe\_8\_3\_5\_100      & 61  & 8  & 40 & 939.33  & 4.32    & \textbf{217.34$\enspace\uparrow$}               & TO      & 50.53   & \textbf{\textgreater{}71.24$\enspace\uparrow$}  \\
		vqe\_8\_4\_10\_100     & 102 & 8  & 71 & TO      & 1037.14 & \textbf{\textgreater{}3.47$\enspace\uparrow$}   & TO      & TO      & N/A -                 \\
		vqe\_8\_4\_5\_100      & 60  & 8  & 39 & 286.91  & 1.34    & \textbf{214.8$\enspace\uparrow$}                & 3472.06 & 31.6    & \textbf{109.88$\enspace\uparrow$}               \\
		Average                &-&-&-&-&-& \textbf{\textgreater{}149.95$\enspace\uparrow$} &-&-& \textbf{\textgreater{}61.21$\enspace\uparrow$}  \\
		\bottomrule
	\end{tabular}
	}
\end{table*}

\begin{table*}[htbp]
	\small
	\tabcolsep=0.008\linewidth
	\renewcommand{\arraystretch}{0.95}
	\caption{Fidelity, Depth, Swap number of HAQA-QSynth-v2 on IBM Eagle.}
	\label{tab:qlt-ibmeagle-qsv2}
	\centering
	\resizebox{\textwidth}{!}{
		\begin{threeparttable}
		\begin{tabular}{ccccccccccr}
			\toprule
			Samples & Depth & $n_q$ & $$ & \multicolumn{2}{c}{Depth} & \multicolumn{2}{c}{Swap Number} & \multicolumn{3}{c}{Fidelity} \\
			\cmidrule(lr){5-6}\cmidrule(lr){7-8}\cmidrule(lr){9-11}
			& & & & QSynth-v2 & HAQA(ours) & QSynth-v2 & HAQA(ours) & QSynth-v2 & HAQA(ours) & \textbf{Percent} \\
			\midrule
			4gt13\_92 & 38 & 5 & 30 & 80 & 78 & 13 & 13 & 0.5732 & 0.6973 & \textbf{21.64\% $\uparrow$} \\
			4mod5-v1\_22 & 12 & 5 & 11 & 22 & 22 & 3 & 3 & 0.8466 & 0.9131 & \textbf{7.85\% $\uparrow$} \\
			adder & 11 & 4 & 10 & 16 & 16 & 2 & 2 & 0.8701 & 0.9004 & \textbf{3.48\% $\uparrow$} \\
			adder\_n10\_transpiled & 119 & 10 & 65 & 162 & 162 & 14 & 14 & 0.5166 & 0.5771 & \textbf{11.72\% $\uparrow$} \\
			barenco\_tof\_4 & 68 & 7 & 34 & 84 & 90 & 8 & 9 & 0.708 & 0.7783 & \textbf{9.93\% $\uparrow$} \\
			barenco\_tof\_5 & 95 & 9 & 50 & 123 & 123 & 14 & 15 & 0.5967 & 0.6357 & \textbf{6.55\% $\uparrow$} \\
			mod5mils\_65 & 21 & 5 & 16 & 45 & 46 & 6 & 6 & 0.7842 & 0.8369 & \textbf{6.73\% $\uparrow$} \\
			or & 8 & 3 & 6 & 18 & 17 & 2 & 2 & 0.9209 & 0.9453 & \textbf{2.65\% $\uparrow$} \\
			qaoa5 & 14 & 5 & 8 & 16 & 15 & 0 & 0 & 0.9829 & 0.9774 & \textbf{0.05\% $\uparrow$} \\
			qpe\_n9\_transpiled & 92 & 9 & 43 & 132 & 123 & 12 & 12 & 0.8206 & 0.8598 & \textbf{4.78\% $\uparrow$} \\
			tof\_4 & 46 & 7 & 22 & 56 & 56 & 3 & 3 & 0.8398 & 0.8516 & \textbf{1.4\% $\uparrow$} \\
			tof\_5 & 61 & 9 & 30 & 73 & 77 & 5 & 5 & 0.789 & 0.7754 & -1.72$\%$ $\downarrow$ \\
			vbe\_adder\_3 & 58 & 10 & 50 & 80 & 80 & 10 & 10 & 0.5596 & 0.6426 & \textbf{14.83\% $\uparrow$} \\
			vqe\_8\_1\_5\_100 & 32 & 6 & 18 & 41 & 44 & 3 & 3 & 0.8613 & 0.8838 & \textbf{2.61\% $\uparrow$} \\
			vqe\_8\_2\_5\_100 & 80 & 8 & 48 & 113 & 112 & 13 & 13 & 0.625 & 0.6533 & \textbf{4.53\% $\uparrow$} \\
			vqe\_8\_3\_5\_100 & 61 & 8 & 40 & 90 & 92 & 10 & 13 & 0.5908 & 0.6748 & \textbf{14.21\% $\uparrow$} \\
			vqe\_8\_4\_5\_100 & 60 & 8 & 39 & 81 & 85 & 8 & 11 & 0.6279 & 0.7266 & \textbf{15.71\% $\uparrow$} \\
			Average &-&-&-&-&-&-&-& 0.7361 & 0.7844 & \textbf{6.52\% $\uparrow$} \\
			\bottomrule
		\end{tabular}
        \begin{tablenotes}
		\footnotesize
		\item As shown in the table, HAQA maintains comparable circuit depth and swap counts while achieving consistent improvements in fidelity compared to the original mapping results.
		\end{tablenotes}
		\end{threeparttable}
	}
\end{table*}

\begin{table*}[htbp]
	\small
	\tabcolsep=0.009\linewidth
	\renewcommand{\arraystretch}{0.95}
	\caption{Fidelity, Depth, Swap number of HAQA-TB-OLSQ2 on IBM Eagle.}
	\label{tab:qlt-ibmeagle-tb2}
	\centering
	\resizebox{\textwidth}{!}{
		\begin{tabular}{ccccccccccr}
			\toprule
			Samples & Depth & $n_q$ & $n_G$ & \multicolumn{2}{c}{Depth} & \multicolumn{2}{c}{Swap Number} & \multicolumn{3}{c}{Fidelity} \\
			\cmidrule(lr){5-6}\cmidrule(lr){7-8}\cmidrule(lr){9-11}
			& & & & TB-OLSQ2 & HAQA(ours) & TB-OLSQ2 & HAQA(ours) & TB-OLSQ2 & HAQA(ours) & \textbf{Percent} \\
			\midrule
			4mod5-v1\_22 & 12 & 5 & 11 & 22 & 23 & 3 & 3 & 0.8701 & 0.9092 & \textbf{4.49\% $\uparrow$} \\
			adder & 11 & 4 & 10 & 16 & 16 & 2 & 2 & 0.9141 & 0.9145 & \textbf{0.04\% $\uparrow$} \\
			barenco\_tof\_4 & 68 & 7 & 34 & 88 & 94 & 8 & 9 & 0.4912 & 0.7871 & \textbf{60.24\% $\uparrow$} \\
			mod5mils\_65 & 21 & 5 & 16 & 49 & 44 & 6 & 6 & 0.708 & 0.8369 & \textbf{18.21\% $\uparrow$} \\
			or & 8 & 3 & 6 & 17 & 15 & 2 & 2 & 0.9394 & 0.96 & \textbf{2.19\% $\uparrow$} \\
			qaoa5 & 14 & 5 & 8 & 15 & 15 & 0 & 0 & 0.9849 & 0.9874 & \textbf{0.25\% $\uparrow$} \\
			tof\_4 & 46 & 7 & 22 & 51 & 56 & 3 & 3 & 0.8574 & 0.8799 & \textbf{2.62\% $\uparrow$} \\
			tof\_5 & 61 & 9 & 30 & 77 & 76 & 5 & 5 & 0.6777 & 0.7998 & \textbf{18.02\% $\uparrow$} \\
			vqe\_8\_1\_5\_100 & 32 & 6 & 18 & 40 & 40 & 3 & 3 & 0.791 & 0.875 & \textbf{10.62\% $\uparrow$} \\
			vqe\_8\_4\_5\_100 & 60 & 8 & 39 & 73 & 85 & 9 & 11 & 0.5742 & 0.7354 & \textbf{28.06\% $\uparrow$} \\
			Average &-&-&-&-&-&-&-& 0.7808 & 0.8685 & \textbf{11.23\% $\uparrow$} \\
			\bottomrule
		\end{tabular}
	}
\end{table*}

\begin{table*}[htbp]
	\small
	\tabcolsep=0.0078\linewidth
	\renewcommand{\arraystretch}{0.95}
	\caption{Fidelity, Depth, Swap number of HAQA-QSynth-v2 on IBM Heron.}
	\label{tab:qlt-ibmheron-qsv2}
	\centering
	\resizebox{\textwidth}{!}{
		\begin{tabular}{ccccccccccr}
			\toprule
			Samples & Depth & $n_q$ & $n_G$ & \multicolumn{2}{c}{Depth} & \multicolumn{2}{c}{Swap Number} & \multicolumn{3}{c}{Fidelity} \\
			\cmidrule(lr){5-6}\cmidrule(lr){7-8}\cmidrule(lr){9-11}
			& & & & QSynth-v2 & HAQA(ours) & QSynth-v2 & HAQA(ours) & QSynth-v2 & HAQA(ours) & \textbf{Percent} \\
			\midrule
			4gt13\_92 & 38 & 5 & 30 & 78 & 77 & 13 & 13 & 0.8174 & 0.872 & \textbf{6.69\% $\uparrow$} \\
			4mod5-v1\_22 & 12 & 5 & 11 & 22 & 22 & 3 & 3 & 0.9385 & 0.968 & \textbf{3.12\% $\uparrow$} \\
			adder & 11 & 4 & 10 & 16 & 16 & 2 & 2 & 0.9053 & 0.978 & \textbf{7.98\% $\uparrow$} \\
			adder\_n10\_transpiled & 119 & 10 & 65 & 159 & 160 & 14 & 14 & 0.7461 & 0.613 & -17.8\% $\downarrow$ \\
			barenco\_tof\_4 & 68 & 7 & 34 & 81 & 87 & 8 & 8 & 0.832 & 0.69 & -17.02\% $\downarrow$ \\
			barenco\_tof\_5 & 95 & 9 & 50 & 123 & 126 & 14 & 14 & 0.6738 & 0.636 & -5.65\% $\downarrow$\\
			mod5mils\_65 & 21 & 5 & 16 & 45 & 45 & 6 & 6 & 0.9072 & 0.947 & \textbf{4.41\% $\uparrow$} \\
			or & 8 & 3 & 6 & 17 & 17 & 2 & 2 & 0.9717 & 0.985 & \textbf{1.41\% $\uparrow$} \\
			qaoa5 & 14 & 5 & 8 & 15 & 15 & 0 & 0 & 0.9839 & 0.992 & \textbf{0.8\% $\uparrow$} \\
			qpe\_n9\_transpiled & 92 & 9 & 43 & 128 & 123 & 12 & 12 & 0.7985 & 0.915 & \textbf{14.61\% $\uparrow$} \\
			tof\_4 & 46 & 7 & 22 & 56 & 56 & 3 & 3 & 0.6543 & 0.833 & \textbf{27.31\% $\uparrow$} \\
			tof\_5 & 61 & 9 & 30 & 77 & 75 & 5 & 5 & 0.9258 & 0.901 & -2.64\% $\downarrow$ \\
			vbe\_adder\_3 & 58 & 10 & 50 & 80 & 80 & 10 & 10 & 0.7852 & 0.762 & -2.99\% $\downarrow$ \\
			vqe\_8\_1\_5\_100 & 32 & 6 & 18 & 44 & 43 & 3 & 3 & 0.5801 & 0.886 & \textbf{52.69\% $\uparrow$} \\
			vqe\_8\_2\_5\_100 & 80 & 8 & 48 & 117 & 114 & 13 & 13 & 0.7959 & 0.845 & \textbf{6.13\% $\uparrow$} \\
			vqe\_8\_3\_5\_100 & 61 & 8 & 40 & 90 & 90 & 10 & 10 & 0.6543 & 0.738 & \textbf{12.84\% $\uparrow$} \\
			vqe\_8\_4\_5\_100 & 60 & 8 & 39 & 82 & 82 & 8 & 8 & 0.8779 & 0.885 & \textbf{0.78\% $\uparrow$} \\
			Average &-&-&-&-&-&-&-& 0.8146 & 0.85 & \textbf{6.69\% $\uparrow$} \\
			\bottomrule
		\end{tabular}
	}
\end{table*}

\begin{table*}[htbp]
	\small
	\tabcolsep=0.0078\linewidth
	\renewcommand{\arraystretch}{1}
	\caption{Fidelity, Depth, Swap number of HAQA-TB-OLSQ2 on IBM Heron.}
	\label{tab:qlt-ibmheron-tb2}
	\centering
	\resizebox{\textwidth}{!}{
		\begin{tabular}{ccccccccccr}
			\toprule
			Samples & Depth & $n_q$ & $n_G$ & \multicolumn{2}{c}{Depth} & \multicolumn{2}{c}{Swap Number} & \multicolumn{3}{c}{Fidelity} \\
			\cmidrule(lr){5-6}\cmidrule(lr){7-8}\cmidrule(lr){9-11}
			& & & & TB-OLSQ2 & HAQA(ours) & TB-OLSQ2 & HAQA(ours) & TB-OLSQ2 & HAQA(ours) & Percent \\
			\midrule
			4mod5-v1\_22 & 12 & 5 & 11 & 22 & 23 & 3 & 3 & 0.8857 & 0.9619 & \textbf{8.6\% $\uparrow$} \\
			adder & 11 & 4 & 10 & 16 & 16 & 2 & 2 & 0.9482 & 0.9219 & -2.78\% $\downarrow$ \\
			barenco\_tof\_4 & 68 & 7 & 34 & 84 & 87 & 8 & 8 & 0.7715 & 0.8994 & \textbf{16.58\% $\uparrow$} \\
			mod5mils\_65 & 21 & 5 & 16 & 47 & 45 & 6 & 6 & 0.2832 & 0.958 & \textbf{238.28\% $\uparrow$} \\
			or & 8 & 3 & 6 & 18 & 16 & 2 & 2 & 0.9736 & 0.9863 & \textbf{1.3\% $\uparrow$} \\
			qaoa5 & 14 & 5 & 8 & 15 & 15 & 0 & 0 & 0.9868 & 0.9884 & \textbf{0.17\% $\uparrow$} \\
			tof\_4 & 46 & 7 & 22 & 52 & 51 & 3 & 3 & 0.916 & 0.7998 & -12.69\% $\downarrow$ \\
			vqe\_8\_1\_5\_100 & 32 & 6 & 18 & 40 & 39 & 3 & 3 & 0.9131 & 0.9609 & \textbf{5.24\% $\uparrow$} \\
			vqe\_8\_4\_5\_100 & 60 & 8 & 39 & 73 & 73 & 9 & 9 & 0.834 & 0.8223 & -1.41\% $\downarrow$ \\
			Average &-&-&-&-&-&-&-& 0.8347 & 0.9218 & \textbf{10.47\% $\uparrow$} \\
			\bottomrule
		\end{tabular}
	}
\end{table*}

\subsection{Fidelity}

We evaluated HAQA's improvements over QSynth-v2 and TB-OLSQ2 on both IBM Eagle and IBM Heron quantum processors. The experimental results are presented in Tables \ref{tab:qlt-ibmeagle-qsv2} through \ref{tab:qlt-ibmheron-tb2}, focusing exclusively on circuits solvable by both QSynth-v2 and TB-OLSQ2 for effective quality metrics comparison.
On IBM Eagle, HAQA-QSynth-v2 achieved up to 21.64\% fidelity improvement over the original QSynth-v2, with an average increase of 6.52\% across all test cases. Compared to the original TB-OLSQ2, fidelity improvements were observed across all samples, reaching up to 60.24\% with an average increase of 11.23\%. Notably, HAQA implementation had minimal impact on circuit depth and swap count.

\begin{figure}[htbp] %
	\centering
	\subfloat[TB-OLSQ2.]{\includegraphics[width=0.24\columnwidth]{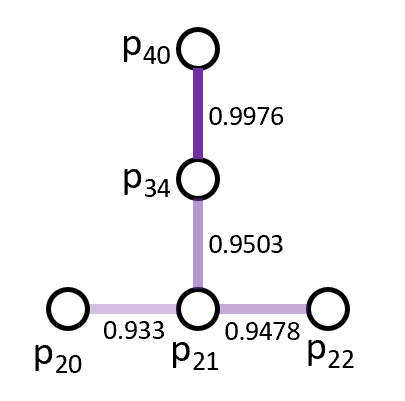}\label{heron-tbolsq2-region}}
	\hspace{0.01\columnwidth}
	\subfloat[HAQA-TB-OLSQ2.]{\includegraphics[width=0.24\columnwidth]{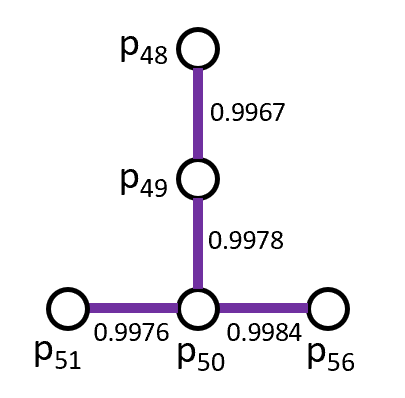}\label{heron-haqa-region}}
	\caption{Mapping regions after TB-OLSQ2 and HAQA-TB-OLSQ2, where darker edges represent higher two-qubit gate fidelity and lighter edges indicate lower two-qubit gate fidelity. The average fidelity of edges on IBM Heron is 0.993.}
	\label{fig:region-compare}
\end{figure}

For IBM Heron, HAQA applied to QSynth-v2 demonstrated a maximum fidelity improvement of 52.69\% with an average increase of 6.69\% across all samples. When applied to TB-OLSQ2, the maximum fidelity improvement reached 238.28\% with an average increase of 10.47\%. A notable example is the $mod5mils\_65$ circuit, where fidelity improved from 0.2832 with TB-OLSQ2 to 0.958 with HAQA. Figure \ref{fig:region-compare} illustrates the operational regions of TB-OLSQ2 and HAQA-TB-OLSQ2 mapped circuits on IBM Heron. While TB-OLSQ2 and HAQA-TB-OLSQ2 mapped to physical qubits $p_{20}$, $p_{21}$, $p_{22}$, $p_{34}$, $p_{40}$ and $p_{48}$, $p_{49}$, $p_{50}$, $p_{51}$, $p_{56}$ respectively with identical topological structures, the two-qubit gate fidelities differ significantly. HAQA's mapping region maintains two-qubit gate fidelities above the average across all edges, effectively avoiding extreme low fidelity scenarios and enhancing qubit mapping stability. 

Based on Tables \ref{tab:qlt-ibmeagle-qsv2} to \ref{tab:qlt-ibmheron-tb2}, comprehensive experimental analysis demonstrates HAQA's effectiveness in enhancing solver-based qubit mapping methods. Working in conjunction with existing solvers, the method maintains comparable performance in critical mapping metrics while achieving hundred-fold acceleration ratios, consistent with theoretical complexity analysis predictions. HAQA effectively prevents extreme low fidelity scenarios through hardware-aware region selection, enhancing mapping stability across different quantum architectures. The experimental results demonstrate that fidelity improvements correlate positively with the number of two-qubit gates in quantum circuits, highlighting HAQA's particular significance for implementing complex quantum circuits. These results demonstrate that HAQA successfully addresses both the efficiency bottleneck and fidelity optimization challenges in solver-based qubit mapping approaches.

\section{Discussion}
HAQA provides an initial approach to address the core challenges in qubit mapping through hardware-guided regional optimization, reveal that quantum programs can achieve higher fidelity while utilizing substantially fewer physical qubits than the quantum computer contains. This strategy shows increasing potential as quantum circuits and architectures scale up, where the efficiency difference between global and regional optimization approaches becomes more pronounced. The quantitative analysis of quantum hardware connectivity patterns' impact on mapping efficiency represents a promising research direction. The modeling and analysis approaches discussed in this work could provide useful references for investigating these architectural considerations.

\section{Conclusion}
This work proposes HAQA, a hardware-aware and adaptive method for efficient qubit mapping. The method addresses both efficiency and fidelity challenges through hardware-guided region identification and adaptive expansion mechanisms, transforming the global mapping problem into guided regional optimization. The computational complexity analysis demonstrates polynomial-level acceleration potential, with the analytical framework offering a transferable approach for evaluating solver-based mapping methods. While maintaining comparable performance in circuit depth and swap count with existing solvers, HAQA achieves significant improvements with acceleration ratios of up to $632.76\times$ and $286.87\times$ for Qsynth-v2 and TB-OLSQ2 respectively, alongside fidelity improvements of up to $52.69\%$ and $238.28\%$. These results demonstrate HAQA's practical applicability as an enhancement to current solver-based approaches. The optimization principles presented in our work are applicable to various solver-based mapping approaches, offering potential improvements for existing and future qubit mapping methodologies.

\section*{Declaration}
\textbf{Availability of data and materials}\\
The code for this work is available at: \url{https://github.com/StillwaterQ/HAQA}.\\
\textbf{Competing interests}\\
The authors declare that they have no competing interests as defined by Springer, or other interests that might be perceived to influence the results and/or discussion reported in this paper.\\
\textbf{Funding}\\
This work was supported in part by the National Natural Science Foundation of China under Grant 62472072, and in part by the National Natural Science Foundation of China under Grant 62172075.\\
\textbf{Authors' contributions}\\
Wenjie Sun was responsible for the conceptualization, formal analysis, methodology design, software implementation, and writing of the original draft.
Xiaoyu Li contributed to funding acquisition.
Lianhui Yu performed data curation and validation.
Zhigang Wang supervised the project.
Geng Chen contributed to formal analysis and visualization.
Desheng Zheng managed the project administration.
Guowu Yang contributed to funding acquisition and provided resources.

\bibliography{mylib}


\begin{thebibliography}{28}
\ifx \bisbn   \undefined \def \bisbn  #1{ISBN #1}\fi
\ifx \binits  \undefined \def \binits#1{#1}\fi
\ifx \bauthor  \undefined \def \bauthor#1{#1}\fi
\ifx \batitle  \undefined \def \batitle#1{#1}\fi
\ifx \bjtitle  \undefined \def \bjtitle#1{#1}\fi
\ifx \bvolume  \undefined \def \bvolume#1{\textbf{#1}}\fi
\ifx \byear  \undefined \def \byear#1{#1}\fi
\ifx \bissue  \undefined \def \bissue#1{#1}\fi
\ifx \bfpage  \undefined \def \bfpage#1{#1}\fi
\ifx \blpage  \undefined \def \blpage #1{#1}\fi
\ifx \burl  \undefined \def \burl#1{\textsf{#1}}\fi
\ifx \doiurl  \undefined \def \doiurl#1{\url{https://doi.org/#1}}\fi
\ifx \betal  \undefined \def \betal{\textit{et al.}}\fi
\ifx \binstitute  \undefined \def \binstitute#1{#1}\fi
\ifx \binstitutionaled  \undefined \def \binstitutionaled#1{#1}\fi
\ifx \bctitle  \undefined \def \bctitle#1{#1}\fi
\ifx \beditor  \undefined \def \beditor#1{#1}\fi
\ifx \bpublisher  \undefined \def \bpublisher#1{#1}\fi
\ifx \bbtitle  \undefined \def \bbtitle#1{#1}\fi
\ifx \bedition  \undefined \def \bedition#1{#1}\fi
\ifx \bseriesno  \undefined \def \bseriesno#1{#1}\fi
\ifx \blocation  \undefined \def \blocation#1{#1}\fi
\ifx \bsertitle  \undefined \def \bsertitle#1{#1}\fi
\ifx \bsnm \undefined \def \bsnm#1{#1}\fi
\ifx \bsuffix \undefined \def \bsuffix#1{#1}\fi
\ifx \bparticle \undefined \def \bparticle#1{#1}\fi
\ifx \barticle \undefined \def \barticle#1{#1}\fi
\bibcommenthead
\ifx \bconfdate \undefined \def \bconfdate #1{#1}\fi
\ifx \botherref \undefined \def \botherref #1{#1}\fi
\ifx \url \undefined \def \url#1{\textsf{#1}}\fi
\ifx \bchapter \undefined \def \bchapter#1{#1}\fi
\ifx \bbook \undefined \def \bbook#1{#1}\fi
\ifx \bcomment \undefined \def \bcomment#1{#1}\fi
\ifx \oauthor \undefined \def \oauthor#1{#1}\fi
\ifx \citeauthoryear \undefined \def \citeauthoryear#1{#1}\fi
\ifx \endbibitem  \undefined \def \endbibitem {}\fi
\ifx \bconflocation  \undefined \def \bconflocation#1{#1}\fi
\ifx \arxivurl  \undefined \def \arxivurl#1{\textsf{#1}}\fi
\csname PreBibitemsHook\endcsname

\bibitem{Arute2019}
\begin{barticle}
\bauthor{\bsnm{Arute}, \binits{F.}}, \betal:
\batitle{Quantum supremacy using a programmable superconducting processor}.
\bjtitle{Nature}
\bvolume{574}(\bissue{7779}),
\bfpage{505}--\blpage{510}
(\byear{2019}).
\doiurl{10.1038/s41586-019-1666-5}.
\bcomment{Number: 7779 Publisher: Nature Publishing Group}
\end{barticle}
\endbibitem

\bibitem{JerryChow}
\begin{botherref}
\oauthor{\bsnm{{Jerry Chow}}},
\oauthor{\bsnm{{Oliver Dial}}},
\oauthor{\bsnm{{Jay Gambetta}}}:
{IBM} {Quantum} breaks the 100‑qubit processor barrier {\textbar} {IBM}
  {Quantum} {Computing} {Blog}.
\url{https://www.ibm.com/quantum/blog/127-qubit-quantum-processor-eagle}
\end{botherref}
\endbibitem

\bibitem{author.fullNamevphantom}
\begin{botherref}
\oauthor{\bsnm{Padavic-Callaghan}, \binits{K.}}:
{IBM}’s '{Condor}' quantum computer has more than 1000 qubits
(2023).
\url{https://www.newscientist.com/article/2405789-ibms-condor-quantum-computer-has-more-than-1000-qubits/}
\end{botherref}
\endbibitem

\bibitem{2024}
\begin{botherref}
{IBM} {Launches} {Its} {Most} {Advanced} {Quantum} {Computers}, {Fueling} {New}
  {Scientific} {Value} and {Progress} towards {Quantum} {Advantage}
(2024).
\url{https://newsroom.ibm.com/2024-11-13-ibm-launches-its-most-advanced-quantum-computers,-fueling-new-scientific-value-and-progress-towards-quantum-advantage}
\end{botherref}
\endbibitem

\bibitem{Rigetti&Co2024}
\begin{botherref}
\oauthor{\bsnm{LLC}, \binits{R..C.}}:
Rigetti {Announces} {Public} {Availability} of {Ankaa}-2 {System} with a 2.5x
  {Performance} {Improvement} {Compared} to {Previous} {QPUs}
(2024)
\end{botherref}
\endbibitem

\bibitem{khatri2019quantum}
\begin{barticle}
\bauthor{\bsnm{Khatri}, \binits{S.}},
\bauthor{\bsnm{LaRose}, \binits{R.}},
\bauthor{\bsnm{Poremba}, \binits{A.}},
\bauthor{\bsnm{Cincio}, \binits{L.}},
\bauthor{\bsnm{Sornborger}, \binits{A.T.}},
\bauthor{\bsnm{Coles}, \binits{P.J.}}:
\batitle{Quantum-assisted quantum compiling}.
\bjtitle{Quantum}
\bvolume{3},
\bfpage{140}
(\byear{2019})
\end{barticle}
\endbibitem

\bibitem{Preskill2018}
\begin{barticle}
\bauthor{\bsnm{Preskill}, \binits{J.}}:
\batitle{Quantum {Computing} in the {NISQ} era and beyond}.
\bjtitle{Quantum}
\bvolume{2},
\bfpage{79}
(\byear{2018}).
\doiurl{10.22331/q-2018-08-06-79}.
\bcomment{Publisher: Verein zur Förderung des Open Access Publizierens in den
  Quantenwissenschaften}
\end{barticle}
\endbibitem

\bibitem{Tan2020}
\begin{bchapter}
\bauthor{\bsnm{Tan}, \binits{B.}},
\bauthor{\bsnm{Cong}, \binits{J.}}:
\bctitle{Optimal {Layout} {Synthesis} for {Quantum} {Computing}}.
In: \bbtitle{2020 {IEEE}/{ACM} {International} {Conference} {On} {Computer}
  {Aided} {Design} ({ICCAD})},
pp. \bfpage{1}--\blpage{9}
(\byear{2020}).
\bcomment{ISSN: 1558-2434}.
\burl{https://ieeexplore.ieee.org/document/9256696}
\end{bchapter}
\endbibitem

\bibitem{Ferrari2022}
\begin{bchapter}
\bauthor{\bsnm{Ferrari}, \binits{D.}},
\bauthor{\bsnm{Amoretti}, \binits{M.}}:
\bctitle{Noise-adaptive quantum compilation strategies evaluated with
  application-motivated benchmarks}.
In: \bbtitle{Proceedings of the 19th {ACM} {International} {Conference} on
  {Computing} {Frontiers}}.
\bsertitle{{CF} '22},
pp. \bfpage{237}--\blpage{243}.
\bpublisher{Association for Computing Machinery},
\blocation{New York, NY, USA}
(\byear{2022}).
\doiurl{10.1145/3528416.3530250}.
\burl{https://doi.org/10.1145/3528416.3530250}
\end{bchapter}
\endbibitem

\bibitem{Li2019}
\begin{bchapter}
\bauthor{\bsnm{Li}, \binits{G.}},
\bauthor{\bsnm{Ding}, \binits{Y.}},
\bauthor{\bsnm{Xie}, \binits{Y.}}:
\bctitle{Tackling the {Qubit} {Mapping} {Problem} for {NISQ}-{Era} {Quantum}
  {Devices}}.
In: \bbtitle{Proceedings of the {Twenty}-{Fourth} {International} {Conference}
  on {Architectural} {Support} for {Programming} {Languages} and {Operating}
  {Systems}}.
\bsertitle{{ASPLOS} '19},
pp. \bfpage{1001}--\blpage{1014}.
\bpublisher{Association for Computing Machinery},
\blocation{New York, NY, USA}
(\byear{2019}).
\doiurl{10.1145/3297858.3304023}.
\burl{https://dl.acm.org/doi/10.1145/3297858.3304023}
\end{bchapter}
\endbibitem

\bibitem{Niu2020}
\begin{barticle}
\bauthor{\bsnm{Niu}, \binits{S.}},
\bauthor{\bsnm{Suau}, \binits{A.}},
\bauthor{\bsnm{Staffelbach}, \binits{G.}},
\bauthor{\bsnm{Todri-Sanial}, \binits{A.}}:
\batitle{A {Hardware}-{Aware} {Heuristic} for the {Qubit} {Mapping} {Problem}
  in the {NISQ} {Era}}.
\bjtitle{IEEE Transactions on Quantum Engineering}
\bvolume{1},
\bfpage{1}--\blpage{14}
(\byear{2020}).
\doiurl{10.1109/TQE.2020.3026544}.
\bcomment{arXiv:2010.03397 [quant-ph]}
\end{barticle}
\endbibitem

\bibitem{Sivarajah2020}
\begin{barticle}
\bauthor{\bsnm{Sivarajah}, \binits{S.}},
\bauthor{\bsnm{Dilkes}, \binits{S.}},
\bauthor{\bsnm{Cowtan}, \binits{A.}},
\bauthor{\bsnm{Simmons}, \binits{W.}},
\bauthor{\bsnm{Edgington}, \binits{A.}},
\bauthor{\bsnm{Duncan}, \binits{R.}}:
\batitle{t{\textbar}ket⟩: a retargetable compiler for {NISQ} devices}.
\bjtitle{Quantum Science and Technology}
\bvolume{6}(\bissue{1}),
\bfpage{014003}
(\byear{2020}).
\doiurl{10.1088/2058-9565/ab8e92}.
\bcomment{Publisher: IOP Publishing}
\end{barticle}
\endbibitem

\bibitem{Wille2014}
\begin{bchapter}
\bauthor{\bsnm{Wille}, \binits{R.}},
\bauthor{\bsnm{Lye}, \binits{A.}},
\bauthor{\bsnm{Drechsler}, \binits{R.}}:
\bctitle{Optimal {SWAP} gate insertion for nearest neighbor quantum circuits}.
In: \bbtitle{2014 19th {Asia} and {South} {Pacific} {Design} {Automation}
  {Conference} ({ASP}-{DAC})},
pp. \bfpage{489}--\blpage{494}
(\byear{2014}).
\doiurl{10.1109/ASPDAC.2014.6742939}.
\bcomment{ISSN: 2153-697X}.
\burl{https://ieeexplore.ieee.org/document/6742939}
\end{bchapter}
\endbibitem

\bibitem{Wille2019}
\begin{bchapter}
\bauthor{\bsnm{Wille}, \binits{R.}},
\bauthor{\bsnm{Burgholzer}, \binits{L.}},
\bauthor{\bsnm{Zulehner}, \binits{A.}}:
\bctitle{Mapping {Quantum} {Circuits} to {IBM} {QX} {Architectures} {Using} the
  {Minimal} {Number} of {SWAP} and {H} {Operations}}.
In: \bbtitle{2019 56th {ACM}/{IEEE} {Design} {Automation} {Conference}
  ({DAC})},
pp. \bfpage{1}--\blpage{6}
(\byear{2019}).
\bcomment{ISSN: 0738-100X}.
\burl{https://ieeexplore.ieee.org/document/8807099}
\end{bchapter}
\endbibitem

\bibitem{Tan2021}
\begin{bchapter}
\bauthor{\bsnm{Tan}, \binits{B.}},
\bauthor{\bsnm{Cong}, \binits{J.}}:
\bctitle{Optimal qubit mapping with simultaneous gate absorption}.
In: \bbtitle{2021 {IEEE}/{ACM} {International} {Conference} {On} {Computer}
  {Aided} {Design} ({ICCAD})},
pp. \bfpage{1}--\blpage{8}
(\byear{2021}).
\doiurl{10.1109/ICCAD51958.2021.9643554}.
\bcomment{ISSN: 1558-2434}.
\burl{https://ieeexplore.ieee.org/document/9643554}
\end{bchapter}
\endbibitem

\bibitem{Lin2023}
\begin{bchapter}
\bauthor{\bsnm{Lin}, \binits{W.-H.}},
\bauthor{\bsnm{Kimko}, \binits{J.}},
\bauthor{\bsnm{Tan}, \binits{B.}},
\bauthor{\bsnm{Bjørner}, \binits{N.}},
\bauthor{\bsnm{Cong}, \binits{J.}}:
\bctitle{Scalable {Optimal} {Layout} {Synthesis} for {NISQ} {Quantum}
  {Processors}}.
In: \bbtitle{2023 60th {ACM}/{IEEE} {Design} {Automation} {Conference}
  ({DAC})},
pp. \bfpage{1}--\blpage{6}
(\byear{2023}).
\doiurl{10.1109/DAC56929.2023.10247760}.
\burl{https://ieeexplore.ieee.org/document/10247760}
\end{bchapter}
\endbibitem

\bibitem{Shaik2023}
\begin{bchapter}
\bauthor{\bsnm{Shaik}, \binits{I.}},
\bauthor{\bparticle{van~de} \bsnm{Pol}, \binits{J.}}:
\bctitle{Optimal {Layout} {Synthesis} for {Quantum} {Circuits} as {Classical}
  {Planning}}.
In: \bbtitle{2023 {IEEE}/{ACM} {International} {Conference} on {Computer}
  {Aided} {Design} ({ICCAD})},
pp. \bfpage{1}--\blpage{9}
(\byear{2023}).
\doiurl{10.1109/ICCAD57390.2023.10323924}.
\bcomment{ISSN: 1558-2434}.
\burl{https://ieeexplore.ieee.org/abstract/document/10323924}
\end{bchapter}
\endbibitem

\bibitem{Shaik2024}
\begin{bchapter}
\bauthor{\bsnm{Shaik}, \binits{I.}},
\bauthor{\bparticle{van~de} \bsnm{Pol}, \binits{J.}}:
\bctitle{Optimal {Layout} {Synthesis} for {Deep} {Quantum} {Circuits} on {NISQ}
  {Processors} with 100+ {Qubits}}.
In: \bbtitle{{DROPS}-{IDN}/v2/document/10.4230/{LIPIcs}.{SAT}.2024.26}
(\byear{2024}).
\doiurl{10.4230/LIPIcs.SAT.2024.26}.
\burl{https://drops.dagstuhl.de/entities/document/10.4230/LIPIcs.SAT.2024.26}
\end{bchapter}
\endbibitem

\bibitem{Xu2020}
\begin{barticle}
\bauthor{\bsnm{Xu}, \binits{Y.}},
\bauthor{\bsnm{Chu}, \binits{J.}},
\bauthor{\bsnm{Yuan}, \binits{J.}},
\bauthor{\bsnm{Qiu}, \binits{J.}},
\bauthor{\bsnm{Zhou}, \binits{Y.}},
\bauthor{\bsnm{Zhang}, \binits{L.}},
\bauthor{\bsnm{Tan}, \binits{X.}},
\bauthor{\bsnm{Yu}, \binits{Y.}},
\bauthor{\bsnm{Liu}, \binits{S.}},
\bauthor{\bsnm{Li}, \binits{J.}},
\bauthor{\bsnm{Yan}, \binits{F.}},
\bauthor{\bsnm{Yu}, \binits{D.}}:
\batitle{High-{Fidelity}, {High}-{Scalability} {Two}-{Qubit} {Gate} {Scheme}
  for {Superconducting} {Qubits}}.
\bjtitle{Physical Review Letters}
\bvolume{125}(\bissue{24}),
\bfpage{240503}
(\byear{2020}).
\doiurl{10.1103/PhysRevLett.125.240503}.
\bcomment{Publisher: American Physical Society}
\end{barticle}
\endbibitem

\bibitem{Reich2013}
\begin{barticle}
\bauthor{\bsnm{Reich}, \binits{D.M.}},
\bauthor{\bsnm{Gualdi}, \binits{G.}},
\bauthor{\bsnm{Koch}, \binits{C.P.}}:
\batitle{Optimal {Strategies} for {Estimating} the {Average} {Fidelity} of
  {Quantum} {Gates}}.
\bjtitle{Physical Review Letters}
\bvolume{111}(\bissue{20}),
\bfpage{200401}
(\byear{2013}).
\doiurl{10.1103/PhysRevLett.111.200401}.
\bcomment{Publisher: American Physical Society}
\end{barticle}
\endbibitem

\bibitem{Tan2021a}
\begin{barticle}
\bauthor{\bsnm{Tan}, \binits{B.}},
\bauthor{\bsnm{Cong}, \binits{J.}}:
\batitle{Optimality {Study} of {Existing} {Quantum} {Computing} {Layout}
  {Synthesis} {Tools}}.
\bjtitle{IEEE Transactions on Computers}
\bvolume{70}(\bissue{9}),
\bfpage{1363}--\blpage{1373}
(\byear{2021}).
\doiurl{10.1109/TC.2020.3009140}.
\bcomment{Conference Name: IEEE Transactions on Computers}
\end{barticle}
\endbibitem

\bibitem{Huang2019}
\begin{barticle}
\bauthor{\bsnm{Huang}, \binits{W.}},
\bauthor{\bsnm{Yang}, \binits{C.H.}},
\bauthor{\bsnm{Chan}, \binits{K.W.}},
\bauthor{\bsnm{Tanttu}, \binits{T.}},
\bauthor{\bsnm{Hensen}, \binits{B.}},
\bauthor{\bsnm{Leon}, \binits{R.C.C.}},
\bauthor{\bsnm{Fogarty}, \binits{M.A.}},
\bauthor{\bsnm{Hwang}, \binits{J.C.C.}},
\bauthor{\bsnm{Hudson}, \binits{F.E.}},
\bauthor{\bsnm{Itoh}, \binits{K.M.}},
\bauthor{\bsnm{Morello}, \binits{A.}},
\bauthor{\bsnm{Laucht}, \binits{A.}},
\bauthor{\bsnm{Dzurak}, \binits{A.S.}}:
\batitle{Fidelity benchmarks for two-qubit gates in silicon}.
\bjtitle{Nature}
\bvolume{569}(\bissue{7757}),
\bfpage{532}--\blpage{536}
(\byear{2019}).
\doiurl{10.1038/s41586-019-1197-0}.
\bcomment{Publisher: Nature Publishing Group}
\end{barticle}
\endbibitem

\bibitem{Liu2020}
\begin{bchapter}
\bauthor{\bsnm{Liu}, \binits{J.}},
\bauthor{\bsnm{Zhou}, \binits{H.}}:
\bctitle{Reliability {Modeling} of {NISQ}- {Era} {Quantum} {Computers}}.
In: \bbtitle{2020 {IEEE} {International} {Symposium} on {Workload}
  {Characterization} ({IISWC})},
pp. \bfpage{94}--\blpage{105}
(\byear{2020}).
\doiurl{10.1109/IISWC50251.2020.00018}.
\burl{https://ieeexplore.ieee.org/abstract/document/9251243}
\end{bchapter}
\endbibitem

\bibitem{Saravanan2023}
\begin{barticle}
\bauthor{\bsnm{Saravanan}, \binits{V.}},
\bauthor{\bsnm{Saeed}, \binits{S.M.}}:
\batitle{Data-{Driven} {Reliability} {Models} of {Quantum} {Circuit}: {From}
  {Traditional} {ML} to {Graph} {Neural} {Network}}.
\bjtitle{IEEE Transactions on Computer-Aided Design of Integrated Circuits and
  Systems}
\bvolume{42}(\bissue{5}),
\bfpage{1477}--\blpage{1489}
(\byear{2023}).
\doiurl{10.1109/TCAD.2022.3202430}.
\bcomment{Conference Name: IEEE Transactions on Computer-Aided Design of
  Integrated Circuits and Systems}
\end{barticle}
\endbibitem

\bibitem{Sundermann2024}
\begin{barticle}
\bauthor{\bsnm{Sundermann}, \binits{C.}},
\bauthor{\bsnm{Raab}, \binits{H.}},
\bauthor{\bsnm{Heß}, \binits{T.}},
\bauthor{\bsnm{Thüm}, \binits{T.}},
\bauthor{\bsnm{Schaefer}, \binits{I.}}:
\batitle{Reusing d-{DNNFs} for {Efficient} {Feature}-{Model} {Counting}}.
\bjtitle{ACM Trans. Softw. Eng. Methodol.}
\bvolume{33}(\bissue{8}),
\bfpage{208}--\blpage{120832}
(\byear{2024}).
\doiurl{10.1145/3680465}
\end{barticle}
\endbibitem

\bibitem{Guo2024}
\begin{bchapter}
\bauthor{\bsnm{Guo}, \binits{Z.-H.}},
\bauthor{\bsnm{Wang}, \binits{T.-C.}}:
\bctitle{{SMT}-{Based} {Layout} {Synthesis} {Approaches} for {Quantum}
  {Circuits}}.
In: \bbtitle{Proceedings of the 2024 {International} {Symposium} on {Physical}
  {Design}}.
\bsertitle{{ISPD} '24},
pp. \bfpage{235}--\blpage{243}.
\bpublisher{Association for Computing Machinery},
\blocation{New York, NY, USA}
(\byear{2024}).
\doiurl{10.1145/3626184.3633316}.
\burl{https://doi.org/10.1145/3626184.3633316}
\end{bchapter}
\endbibitem

\bibitem{Newman2004}
\begin{botherref}
\oauthor{\bsnm{Newman}, \binits{M.E.J.}}:
Fast algorithm for detecting community structure in networks.
Physical Review E
\textbf{69}(6)
(2004).
\doiurl{10.1103/PhysRevE.69.066133}
\end{botherref}
\endbibitem

\bibitem{qiskit}
\begin{botherref}
Qiskit {\textbar} {IBM} {Quantum} {Computing}.
\url{https://www.ibm.com/quantum/qiskit}
\end{botherref}
\endbibitem

\end{thebibliography}

\begin{appendices}
 	\section{Coupling Graphs}\label{apdix}
	
	\begin{figure}[H] 
		\centering
		\subfloat[IBM Eagle.]{\includegraphics[width=0.7\columnwidth]{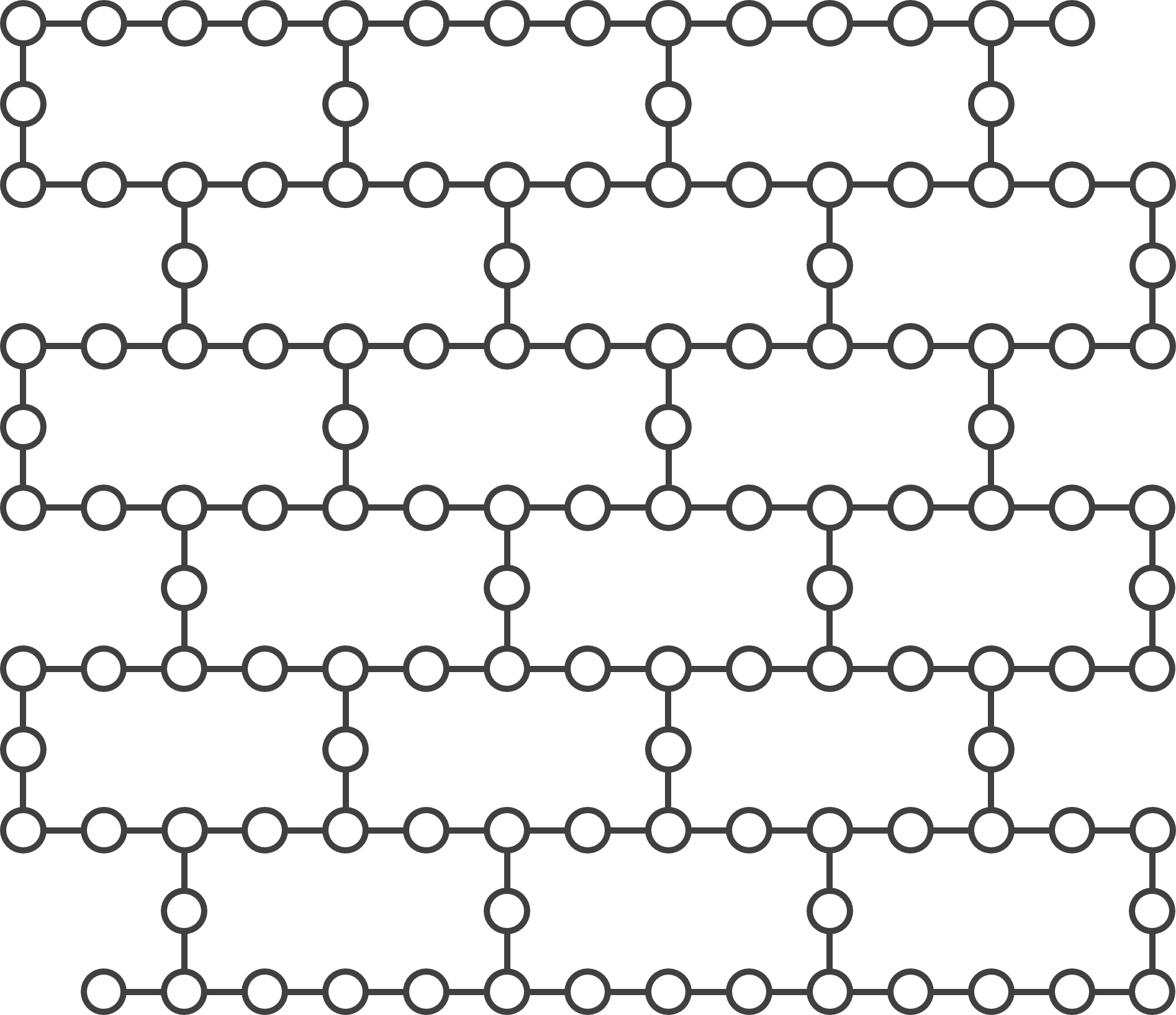}\label{ibm-eagler3-brisbine}} \\
		\vspace{1mm}
		\subfloat[IBM Heron.]{\includegraphics[width=0.7\columnwidth]{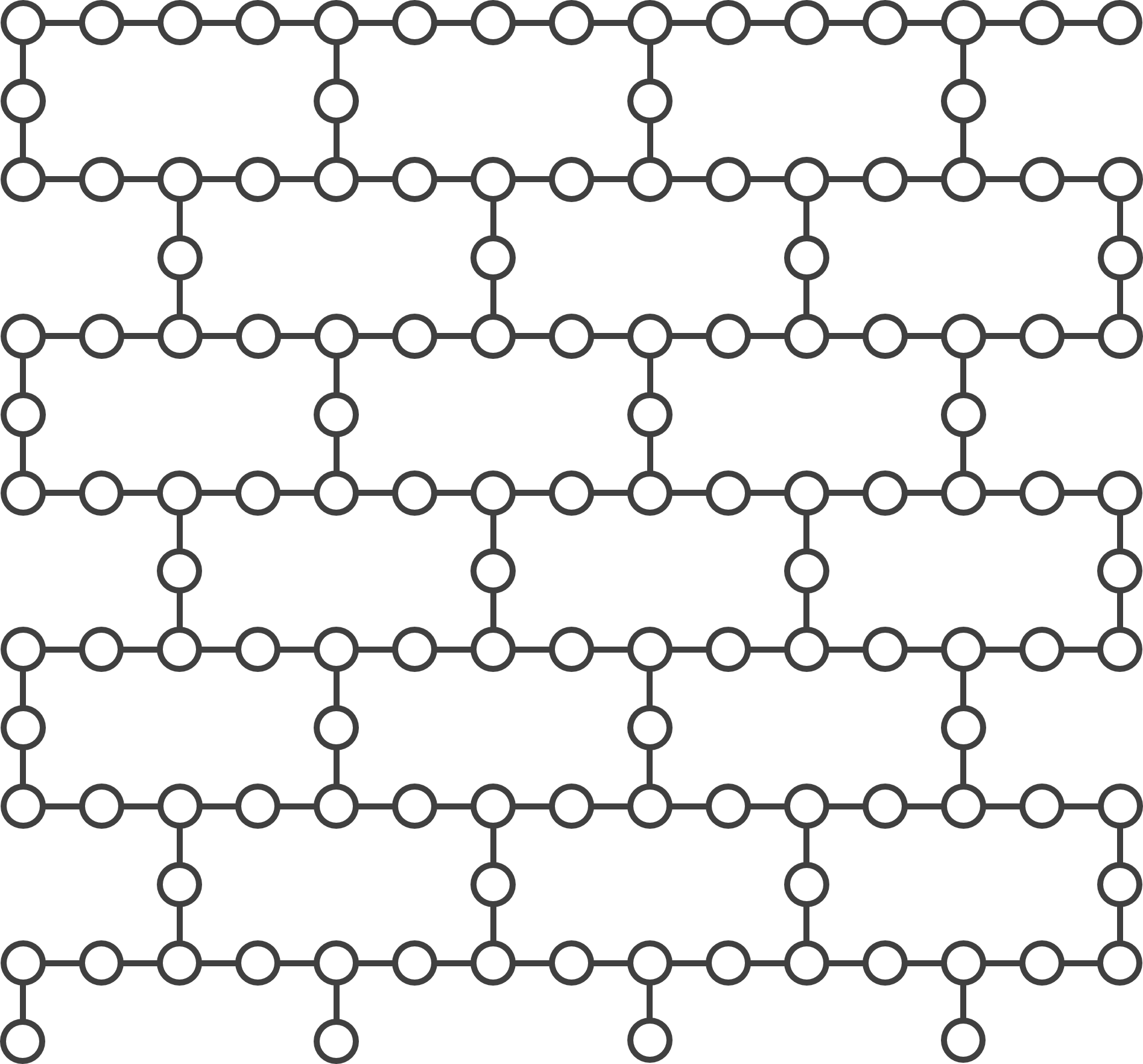}\label{ibm-heronr1-torino}}
		\caption{The coupling graphs used in the experiment, where the circles represent physical qubits and the connecting lines indicate possible two-qubit gates between them. IBM Eagle and IBM Heron have $127$ and $133$ qubits, respectively.}
		\label{exp_coupling}
	\end{figure}
\end{appendices}

\end{document}